\documentclass[%
reprint,
showkeys,
superscriptaddress,
preprintnumbers,
nofootinbib,
amsmath,amssymb,
aps,
prd,
floatfix,
]{revtex4-2}
\usepackage{afterpage}
\usepackage{rotating} 
\usepackage{booktabs}
\usepackage{graphicx}
\usepackage{dcolumn}
\usepackage{bm}
\usepackage{float}
\usepackage{amsmath}
\usepackage{natbib}
\usepackage{hyperref}

\usepackage{xcolor}

\hypersetup{
    colorlinks = true,
    allcolors = blue,
}

\begin{document}

\title{Probing Hidden Dimensions with Multimessenger Gravitational Waves}

\author{Hanlin Song}
\email[Contact author:]{songhanlin@imech.ac.cn}
\affiliation{Center for Gravitational Wave Experiment, National Microgravity Laboratory, Institute of Mechanics, Chinese Academy of Sciences, Beijing 100190, China}

\author{Minghui Du}
\email[Contact author:]{duminghui@imech.ac.cn}
\affiliation{Center for Gravitational Wave Experiment, National Microgravity Laboratory, Institute of Mechanics, Chinese Academy of Sciences, Beijing 100190, China}

\author{Peng Xu}
\email[Contact author:]{xupeng@imech.ac.cn}
\affiliation{Center for Gravitational Wave Experiment, National Microgravity Laboratory, Institute of Mechanics, Chinese Academy of Sciences, Beijing 100190, China}
 \affiliation{Key Laboratory of Gravitational Wave Precision Measurement of Zhejiang Province, Hangzhou Institute for Advanced Study, UCAS, Hangzhou 310024, China}
 \affiliation{Taiji Laboratory for Gravitational Wave Universe (Beijing/Hangzhou), University of Chinese Academy of Sciences (UCAS), Beijing 100049, China}
 \affiliation{Lanzhou Center of Theoretical Physics, Lanzhou University, Lanzhou 730000, China}

\begin{abstract}
The dimensionality of spacetime is a fundamental property of nature, yet whether our observed four-dimensional spacetime is embedded in a higher-dimensional geometry remains experimentally unresolved. Gravitational waves (GWs) offer a direct probe of compact extra dimensions through the strong-field dynamics of binary coalescences. In this work, we investigate extra-dimensional modifications to GW generation within a compactified Kaluza--Klein framework. We derive the Newtonian conservative binary dynamics and the leading-order extra-dimensional correction to the GW energy flux and formulate the corresponding waveform modifications within the parameterized post-Einsteinian (ppE) framework. Combining these results with the previously derived 1PN amplitude correction arising from perturbations of the internal volume, we construct modified GW waveform templates for ground- and space-based detectors. To alleviate parameter degeneracies in GW parameter estimation, we combine the observed strain data of GW170817 with independent electromagnetic observations in a multimessenger Bayesian analysis, yielding a 95\% credible upper bound of $n<0.11$ on the phenomenological dimensional parameter. We further investigate injected multimessenger massive black hole binary observations with Taiji, finding that the upper bound could be improved by nearly two orders of magnitude to $n<0.00178$. These results highlight the potential of multimessenger GW observations to probe extra-dimensional effects and test fundamental gravity in the strong-field and dynamical regime.
\end{abstract}

\maketitle

\section{Introduction}

Whether spacetime contains compact extra spatial dimensions remains an open question in fundamental physics.
Since the proposals of Kaluza and Klein in the 1920s~\cite{Kaluza:1921tu,Klein:1926tv}, Kaluza--Klein (KK) theory has provided one of the most natural geometric extensions of general relativity to higher dimensions. 
In its simplest realization, KK theory begins with Einstein gravity in a higher-dimensional spacetime, from which additional four-dimensional fields emerge upon compactification~\cite{Overduin:1997sri, Bailin:1987jd}.
It established a geometric connection between compact extra dimensions and internal symmetries, providing a conceptual framework that subsequently became central to higher-dimensional supergravity~\cite{Witten:1981me, Duff:1986hr} and string compactifications~\cite{Candelas:1985en}.

More generally, in a $(4+n)$-dimensional spacetime of the form $\mathcal{M}^4\times\mathcal{M}^n$, the extra dimensions constitute a compact internal manifold $\mathcal{M}^n$ with characteristic size $R$.
At energies $E\ll R^{-1}$, only the massless zero modes are dynamically accessible, whereas the massive excitations cannot be produced, rendering the low-energy world effectively four-dimensional. Nevertheless, the geometry of $\mathcal{M}^n$ remains encoded in the field content and interactions of the resulting four-dimensional effective theory~\cite{Duff:1986hr, Bailin:1987jd}. 
Tests of KK gravity therefore provide a direct probe of compact extra dimensions and of the dimensional structure of spacetime itself~\cite{Overduin:1997sri,Duff:1986hr}. 
To date, no compelling evidence for extra spatial dimensions has been found, while existing experimental constraints depend sensitively on the underlying compactification scenario~\cite{ParticleDataGroup:2024cfk}.

In the quasistatic weak-field regime, signatures of extra dimensions have been extensively sought through precision tests of the gravitational inverse-square law at short distances, where hidden dimensions may induce departures from Newtonian gravity. Laboratory torsion-balance experiments have probed gravity from millimeter scales down to separations of $52~\mu\mathrm{m}$, finding no statistically significant deviation from the Newtonian prediction~\cite{Yang:2012zzb,Lee:2020zjt}. Complementary collider and Solar System observations have placed further constraints on model-dependent manifestations of extra dimensions~\cite{Overduin:1997sri,Maartens:2010ar,Clifton:2011jh}. Beyond these conventional tests, gravitational-wave (GW) observations offer a complementary avenue for probing extra-dimensional effects in dynamical gravitational phenomena.

Since the first direct detection of GWs from GW150914~\cite{LIGOScientific:2016aoc}, hundreds of compact-binary mergers have been observed~\cite{LIGOScientific:2026sit}, including binary black holes (BBHs), binary neutron stars (BNSs), and neutron star-black hole binaries (NSBHs). These observations provide new opportunities to investigate extra-dimensional effects on both GW propagation and generation. For example, gravitational radiation may leak into extra dimensions during propagation, modifying the standard inverse-distance scaling of the observed GW amplitude~\cite{Deffayet:2007kf}. This phenomenological effect has been constrained using GW170817~\cite{Pardo:2018ipy} and subsequently the GWTC-3~\cite{MaganaHernandez:2021zyc} and GWTC-4.0~\cite{Chen:2026owi} catalogs, with the results remaining consistent with four-dimensional GW propagation. Other studies have explored differences between GW and electromagnetic propagation times~\cite{Lin:2023zna} and additional GW polarization modes arising from extra spatial dimensions~\cite{Wang:2026sir}. Beyond these propagation and polarization effects, compactified KK theories can also modify GW generation through corrections to binary dynamics and gravitational radiation~\cite{Qiang:2016chy,Du:2020rlx}.

In this work, we investigate extra-dimensional modifications to GW generation within the compactified KK framework. We derive the Newtonian conservative dynamics and the leading-order extra-dimensional correction to the GW energy flux, and formulate the resulting 0PN amplitude and phase modifications within the parameterized post-Einsteinian (ppE) framework. Combining these results with the previously derived 1PN amplitude correction arising from perturbations of the compact internal volume~\cite{Qiang:2016chy}, we construct modified GW waveform templates characterized by the dimensionless parameter $\kappa=n/(2+n)$, where $n$ is treated as a continuous phenomenological dimensional parameter.

However, these corrections introduce strong parameter degeneracies during data analysis, particularly among $\kappa$, the luminosity distance $D_L$, and the chirp mass $\mathcal{M}_c$. This motivates the use of multimessenger GW observations, in which independent electromagnetic measurements provide additional information on the source distance and sky position, helping to alleviate these degeneracies. We first analyze the observed strain data of GW170817~\cite{LIGOScientific:2017vwq}, incorporating information from its electromagnetic counterpart~\cite{LIGOScientific:2017zic,LIGOScientific:2017ync,DES:2017kbs,Cantiello:2018ffy,Mooley:2017enz,Mooley:2018qfh,Mooley:2022uqa}, and implement the ppE corrections using the inspiral-only \textsc{TaylorF2} waveform to constrain $\kappa$ through Bayesian inference. We further investigate the prospects for future space-based GW detectors through simulated multimessenger MBHB observations with Taiji~\cite{Hu:2017mde,Du:2025xdq,Mangiagli:2022niy}. To account for higher-order harmonics and the full inspiral--merger--ringdown (IMR) evolution of these high-SNR signals, we adopt \textsc{IMRPhenomHM} as the baseline waveform, extend the ppE corrections from the dominant $(2,2)$ mode to higher-order $(\ell,m)$ modes, and apply Planck-taper windows to both the amplitude and phase corrections to confine their effects to the inspiral regime.

The remainder of this paper is organized as follows. In Sec.~\ref{ppE_waveform_section}, we derive the leading-order extra-dimensional corrections to binary dynamics and GW emission in the KK theory, incorporate the 1PN amplitude correction arising from internal volume perturbations, and construct the corresponding waveform within the ppE framework. In Sec.~\ref{multimessenger_events}, we introduce the two multimessenger GW scenarios considered in this work, including the observed binary neutron star merger GW170817 and simulated massive-black-hole binary observations with Taiji. In Sec.~\ref{extra_dimensional_templates}, we describe the implementation of the modified waveform templates for ground- and space-based GW detectors and present the Bayesian parameter-estimation framework. Finally, in Sec.~\ref{results}, we report the constraints obtained from GW170817, investigate the prospective sensitivity of Taiji to extra-dimensional effects, and discuss the implications of these results. Throughout this work, the Greek indices $\mu,\nu=0,\ldots,3$ label the four-dimensional spacetime, lowercase Latin indices $a,b$ label the compact internal $n$ dimensions, and uppercase Latin indices $M,N=0,\ldots,3+n$ label the full $(4+n)$-dimensional spacetime with signature $(-,+,\ldots,+)$.

\section{Extra-Dimensional Corrections to GW Waveforms in Kaluza–Klein Theory}
\label{ppE_waveform_section}

In this section, we investigate GW  emission from a compact binary whose orbital motion perturbs the $(4+n)$-dimensional spacetime. Dimensional reduction over the compact extra dimensions yields the effective four-dimensional field equations and the corresponding modifications to the GR waveform. Motivated by Ref.~\cite{Du:2020rlx}, we consider a flat internal $n$-torus, $T^n$, and restrict our analysis to the massless KK sector. We derive the Newtonian conservative binary dynamics and leading-order (0PN) corrections to the GW energy flux, and obtain the consistent results with Ref.~\cite{Du:2020rlx}. Furthermore, we account for perturbations of the internal volume, as discussed in Ref.~\cite{Qiang:2016chy}, which modify the effective four-dimensional stress-energy tensor and hence the source of gravitational radiation. This effect introduces an additional 1PN correction to the GW amplitude. Combining these two effects, we construct the GW waveform within the parametrized post-Einsteinian (ppE) framework~\cite{Yunes:2009ke}, which incorporates extra-dimensional corrections to both the GW amplitude and phase in Kaluza–Klein theory.

\subsection{Leading-Order Corrections to Conservative Dynamics and GW Energy Flux}

The action and field equations of the $(4+n)$-dimensional KK theory are described by the Einstein--Hilbert action~\cite{Overduin:1997sri,Qiang:2016chy},
\begin{gather}
    S_D=\frac{1}{16\pi G_D}\int d^{D}X\sqrt{-g_D}\,R_D+S_m,
    \nonumber\\
    R_{MN}-\frac12g_{MN}R_D=8\pi G_D T_{MN},
\end{gather}
where $G_D$ denotes the gravitational coupling constant in $D=4+n$ spacetime dimensions.  Inspired by Ref.~\cite{Du:2020rlx}, we consider a $D$-dimensional spacetime of the form $\mathcal{M}^4\times T^n$ with a flat internal torus~\cite{Eingorn:2010xp,Chopovsky:2013dg}. We adopt the following metric ansatz which allows for isotropic variations of the internal volume,
\begin{align}
    ds_D^2&=g_{\mu\nu}(x)dx^\mu dx^\nu
    +e^{2\sigma(x)}\delta_{ab}dy^a dy^b,
\end{align}
where $x^\mu=(t,x^1,x^2,x^3)$ and $y^a=(y^1,\ldots,y^n)$ denote the coordinates of the four-dimensional spacetime and the compact internal space, respectively. The metrics $g_{\mu\nu}(x)$ and $\gamma_{ab}(x)=e^{2\sigma(x)}\delta_{ab}$ describe the corresponding geometries. To account for variations of the internal volume induced by the binary's gravitational field, we introduce a scalar field $\sigma(x)$ that allows the compactification scale to vary over four-dimensional spacetime~\cite{Qiang:2016chy,Lehners:2022xds}.  Considering $\det\gamma_{ab}=e^{2n\sigma(x)}$, the volume of the compact internal space is given by
\begin{align}
    V(x)&=\int_{T^n}d^ny\sqrt{\gamma}=V_0e^{n\sigma(x)},
    \label{KK_volume}
\end{align}
where $V_0$ denotes the unperturbed volume of the internal torus. Since $\sigma(x)$ is independent of the internal coordinates $y^a$, it represents the KK volume zero mode. For the flat-torus compactification considered here, this mode remains massless in the absence of an additional stabilization potential~\cite{Lehners:2022xds, Overduin:1997sri}.

We consider a nonspinning compact binary undergoing an adiabatic, quasicircular inspiral. Each component is modeled as a point mass in the three noncompact spatial dimensions, uniformly smeared over the internal torus. The corresponding matter action is
\begin{equation}
\begin{aligned}
    S_m &=-\sum_A m_A\int_{T^n}\frac{d^ny}{V_0}\int ds_A \\
        &=-\sum_A m_A\int ds_A.
\end{aligned}
\end{equation}
where $d^ny/V_0$ describes the uniform distribution of each compact object of mass $m_A$ over the internal torus, and $ds_A$ denotes the proper-time element along its four-dimensional worldline. Meanwhile, the energy-momentum tensor take the form,
\begin{equation}
    T_D^{\mu\nu}=\frac{T^{\mu\nu}}{V(x)},
    \qquad T_D^{ab}=T_D^{\mu a}=0.
    \label{eq:KK_source}
\end{equation}
where $T_D^{\mu\nu}$ denotes the external spacetime components of the higher-dimensional stress-energy tensor, while $T^{\mu\nu}$ is the standard four-dimensional stress-energy tensor of the binary. Consequently, the higher-dimensional source density scales inversely with the internal volume $V(x)$, while integration over the internal space recovers $T^{\mu\nu}$. The vanishing components $T_D^{\mu a}=T_D^{ab}=0$ correspond to the absence of internal momentum flux and tension.

To determine the leading-order effects of the compact extra dimensions, we linearize the higher-dimensional metric around the flat background $\widetilde g_{MN}=\operatorname{diag}(\eta_{\mu\nu},\delta_{ab})$,
\begin{equation}
g_{MN}=\widetilde g_{MN}+h_{MN}.
\end{equation}
We introduce the $D$-dimensional trace-reversed metric perturbation and the corresponding inverse relation
\begin{align}
\bar h_{MN}&=h_{MN}-\frac12\widetilde g_{MN}h_D,
\nonumber\\
h_{MN}&=\bar h_{MN}-\frac{\widetilde g_{MN}}{n+2}\bar h_D,
\end{align}
where $h_D=\widetilde g^{MN}h_{MN}$ and $\bar h_D=\widetilde g^{MN}\bar h_{MN}$ denote the corresponding $D$-dimensional traces. In the harmonic gauge, the linearized Einstein equations take the form
\begin{align}
\Box_D\bar h_{MN}&=-16\pi G_D T_{MN},
\qquad \partial^M\bar h_{MN}=0.
\label{EOM}
\end{align}
Since the matter source has no internal momentum or stress, $T_{\mu a}=T_{ab}=0$, the retarded solution with vanishing homogeneous contributions satisfies $\bar h_{\mu a}=\bar h_{ab}=0$. Nevertheless, the internal metric remains perturbed through the trace-reversal relation, which gives
\begin{align}
h_{ab}&=-\frac{\delta_{ab}}{n+2}\bar h^\mu_\mu,
\end{align}
Thus we have
\begin{align}
\sigma(x)=-\frac{1}{2(n+2)}\bar h^\mu_\mu.
\end{align}
Consequently, according to Eq.~\ref{KK_volume}, the fractional variation of the internal volume is
\begin{align}
\frac{\delta V}{V_0}
&=n\sigma(x)
=-\frac{\kappa}{2}\bar h^\mu{}_\mu,
\qquad
\kappa=\frac{n}{n+2}.
\end{align}
The parameter $\kappa$ characterizes the coupling between internal volume perturbations and the four-dimensional gravitational field~\cite{Qiang:2016chy}.

The same trace-reversal relation also modifies the effective Newtonian gravitational coupling. Defining $G \equiv G_D/V_0$, the leading-order solution of Eq.~\ref{EOM} for a static point mass $m$ is
\begin{align}
    \bar h_{00}=4Gm/r, \ \bar h_{ij}=0. 
\end{align}
The corresponding four-dimensional metric perturbations are
\begin{align}
h_{00}=\frac{2Gm}{r}(1+\kappa),
\nonumber\\
h_{ij}=\frac{2Gm}{r}(1-\kappa)\delta_{ij}.
\end{align}
Comparison with the Newtonian limit, $g_{00}=-1+2G_Nm/r$, then yields
\begin{align}
G_N=(1+\kappa)G=(1+\kappa)\frac{G_D}{V_0}.
\end{align}
Thus, the measured Newton constant differs from the dimensionally reduced gravitational coupling due to the extra-dimensional scalar degree of freedom. The limit $\kappa=0$ recovers GR, while values of $n$ correspond to the number of compact spatial dimensions. We also allow $0\leq\kappa<1$ as a continuous phenomenological extension.

For a compact binary with component masses $m_1$ and $m_2$, we can define the total mass $M=m_1+m_2$, reduced mass $\mu=m_1m_2/M$, symmetric mass ratio $\eta=\mu/M$, and chirp mass $\mathcal{M}_c=\eta^{3/5}M$. In terms of the measured Newton constant $G_N$, the Newtonian circular-orbit relations retain their GR form,
\begin{gather}
\Omega^2=\frac{G_NM}{r^3},\\
E_N(x)=-\frac12\mu c^2x,
\end{gather}
where $x\equiv\left({G_NM\Omega}\right)^{2/3}$. 
Thus, the Newtonian binding energy receives no additional extra-dimensional correction.

The leading quadrupole radiation is sourced by both matter and gravitational field stresses. For a Newtonian binary with relative position $r_i=rn_i$ and velocity $v_i$, the conserved effective source gives
\begin{equation}
    \frac12\ddot I_{ij}
    =\mu v_iv_j-\frac{G_N\mu M}{r}n_in_j,
    \qquad I_{ij}=\mu r_i r_j.
\end{equation}
The kinetic and gravitational terms enter at the same Newtonian order. The transverse-traceless metric perturbation at a distance $d$ is given by
\begin{equation}
    h_{ij}^{\rm TT}=2G\ddot I_{ij}^{\rm TT}/(dc^4).
\end{equation}

The total luminosity also includes radiation from the internal-volume zero mode. For a circular binary with sources uniformly smeared over the compact dimensions, we write $S_{ij}=I_{\langle ij\rangle}^{(3)}$ and $S_{ia}=S_{ab}=0$, where angular brackets on indices denote the symmetric trace-free part. The radiative zero modes are projected using the high-dimensional transverse-traceless operator
\begin{equation}
    \Lambda_{IJ,KL}
    =P_{IK}P_{JL}-\frac{1}{n+2}P_{IJ}P_{KL},
    \quad P_{IJ}=\delta_{IJ}-N_IN_J,
    \label{eq:KK_TT_projector}
\end{equation}
where $I,J,K,L$ span all spatial dimensions and $N_I$ points along the noncompact directions. Integrating the energy flux over $S^2\times T^n$ gives
\begin{equation}
\begin{aligned}
    \mathcal{F}_N
    &=\frac{G}{8\pi c^5}\int d\Omega_{\boldsymbol N}
    \left\langle|\Lambda S|^2\right\rangle
    \\
    &=\frac{G}{5c^5}
    \left(1+\frac{\kappa}{6}\right)
    \left\langle S_{ij}S_{ij}\right\rangle,
    \label{eq:KK_flux_projection}
\end{aligned} 
\end{equation}
where angular brackets denote an orbital average. The $\kappa/6$ term accounts for volume-mode quadrupole radiation. Using $\langle S_{ij}S_{ij}\rangle =32\mu^2r^4\Omega^6$, we obtain
\begin{align}
    \mathcal{F}_N
    &=\frac{32}{5}\frac{c^5}{G_N}
    \eta^2x^5\mathcal{K}(\kappa),    
    \label{eq:KK_flux_ratio}
\end{align}
where $\mathcal{K}(\kappa) \equiv{(1+\kappa/6)}/{(1+\kappa)}$.
These results are consistent with Ref.~\cite{Du:2020rlx} at the zero-mode level. This agreement provides a nontrivial consistency check of our leading-order radiation calculation.

\subsection{1PN Amplitude Corrections from Internal Volume Perturbations}

As shown in Ref.~\cite{Qiang:2016chy}, perturbations of the compact internal volume also modify the effective four-dimensional source of gravitational radiation. The retarded Green's function for the effective four-dimensional linearized field equations yields
\begin{equation}
\bar h_{\mu\nu}(t,\boldsymbol{x})
=\frac{4G_N}{(1+\kappa)c^4}
\int\frac{d^3x'}{\mathcal R}
\int_{T^n}d^ny\sqrt{\widetilde\gamma}\,
T^D_{\mu\nu}(t_{\rm ret},\boldsymbol{x}',y),
\label{eq:KK_retarded_solution}
\end{equation}
where $\mathcal R=|\boldsymbol{x}-\boldsymbol{x}'|$
and $t_{\rm ret}=t-\mathcal R/c$.

The Green's function involves the background internal measure $d^ny\sqrt{\widetilde\gamma}$, whereas the physical four-dimensional stress-energy tensor is defined using the perturbed measure $d^ny\sqrt{\gamma}$.
Thus, using $T^D_{\mu\nu}=T_{\mu\nu}/V(x)$, the effective source entering the retarded solution becomes
\begin{align}
T_{\mu\nu}^{\rm src}
&=\int_{T^n}d^ny\sqrt{\widetilde\gamma}\,
T^D_{\mu\nu}
=\frac{V_0}{V(x)}T_{\mu\nu}
\nonumber\\
&=\left(1-\frac{2\kappa}{1+\kappa}
\frac{U_N}{c^2}\right)T_{\mu\nu}
+\mathcal O(c^{-4}),
\end{align}
where $U_N$ is the Newtonian gravitational potential, with $\bar h^\lambda{}_\lambda=-4U_N/[(1+\kappa)c^2]$ at Newtonian order.

Substituting the corrected source into the retarded solution yields the compact internal volume induced contribution
\begin{equation}
\bar h_{\mu\nu}^{\rm vol}
=-\frac{8G_N\kappa}{(1+\kappa)^2c^6}
\int d^3x'\,
\frac{
U_N(t_{\rm ret},\boldsymbol{x}')
T_{\mu\nu}(t_{\rm ret},\boldsymbol{x}')
}{\mathcal R}.
\end{equation}
Since $U_N/c^2\sim v^2/c^2$, this contribution enters at relative 1PN order.

For a quasicircular binary, the corresponding time-domain corrections to the GW polarizations can be written as~\cite{Qiang:2016chy}
\begin{align}
h_+^{\rm vol}(t)
&=-8\kappa x A_{\rm GR}
\frac{1+\cos^2\iota}{2}
\cos\Psi_{\rm GR}(t),
\nonumber\\
h_\times^{\rm vol}(t)
&=-8\kappa x A_{\rm GR}
\cos\iota\sin\Psi_{\rm GR}(t),
\end{align}
where $A_{\rm GR}$ and $\Psi_{\rm GR}(t)$ denote the Newtonian order amplitude and GW phase, respectively. The additional factor of $x$ identifies the volume-induced contribution as a relative 1PN amplitude correction.

\subsection{GW Waveform Corrections in the Parameterized Post-Einsteinian Framework}
\label{ppE_waveform}
To capture the effects of extra dimensions on GW waveforms, we adopt the widely used parameterized post-Einsteinian (ppE) framework, originally proposed by Yunes and Pretorius~\cite{Yunes:2009ke} and widely used to test modified theories of gravity using GW observations~\cite{Cornish:2011ys, Tahura:2018zuq, Song:2025dbf, Zhu:2026wtl}. 
Within this framework, deviations from GR in the GW signal emitted by an inspiraling compact binary can be parameterized as the frequency-domain waveform~\cite{Yunes:2009ke},
\begin{equation}
\tilde{h}(f) = \tilde{h}_{\mathrm{GR}}(f) \left(1 + \alpha u^a \right) e^{i \beta u^b},
\end{equation}
where $\tilde{h}_{\mathrm{GR}}$ denotes the GR waveform, and $\alpha u^a$ and $\delta\Psi = \beta u^b$ characterize the extra-dimensional corrections to the waveform amplitude and phase, respectively. These deviations are parameterized by the four ppE parameters $\{\alpha, \beta, a, b\}$. Similarly, deviations from GR in the binary orbital dynamics can be parameterized as
\begin{align}
    r &= r_{\mathrm{GR}} \left(1 + \gamma_r u^{c_f} \right), \label{orbital2} \\
    \dot{f} &= \dot{f}_{\mathrm{GR}} \left(1 + \gamma_{\dot{f}} u^{c_{\dot{f}}} \right). 
\end{align}

We now map the extra-dimensional corrections derived above onto the GR waveform within the ppE framework. For the dominant quadrupole harmonic, $f=\Omega/\pi$, we introduce the dimensionless frequency parameter
\begin{equation}
u\equiv\left(\frac{\pi G_N\mathcal{M}_c f}{c^3}\right)^{1/3},
\qquad
x=\eta^{-2/5}u^2.
\end{equation}
Since the Newtonian Kepler relation retains its GR form when expressed in terms of $G_N$, the leading conservative correction vanishes, $\gamma_r=0$. In contrast, the modified GW energy flux changes the orbital frequency evolution. Using the energy-balance equation, $dE_N/dt=-\mathcal{F}_N$, we obtain
\begin{equation}
\dot f=\mathcal{K}\dot f_{\rm GR},
\qquad
\mathcal{K}\equiv\frac{1+\kappa/6}{1+\kappa},
\label{eq:KK_frequency_evolution}
\end{equation}
where
\begin{equation}
\dot f_{\rm GR}
=\frac{96}{5}\pi^{8/3}
\left(\frac{G_N\mathcal{M}_c}{c^3}\right)^{5/3}
f^{11/3}.
\end{equation}
The corresponding ppE parameters are therefore
$\gamma_{\dot f}=\mathcal{K}-1$ and
$c_{\dot f}=0$.

Under the stationary phase approximation (SPA) method, the Fourier phase satisfies $d^2\Psi/df^2=2\pi/\dot f$. Integrating this relation yields
\begin{equation}
\Psi_{\rm KK}(f)
=\Psi_{\rm GR}(f)+\beta_0u^{-5},
\label{eq:KK_phase_0PN}
\end{equation}
where
\begin{equation}
\begin{aligned}
\Psi_{\rm GR}(f)
&=2\pi ft_c-\phi_c-\frac{\pi}{4}
+\frac{3}{128}u^{-5},
\\
\beta_0
&=\frac{15\kappa}{128(6+\kappa)}.
\end{aligned}
\end{equation}
Here $t_c$ and $\phi_c$ denote the coalescence time and GW phase, respectively. The correction has the ppE exponent $b_0=-5$, corresponding to a relative 0PN modification of the GW phase. 
The fractional correction to the Newtonian GR phase coefficient is $\delta\hat{\phi}=\beta_0/(3/128)=5\kappa/(6+\kappa)$. For the five-dimensional limit considered by Ref.~\cite{Du:2020rlx}, setting $n=1$ ($\kappa=1/3$) yields $\delta\hat{\phi}=5/19$, in agreement with their leading-order result. This agreement further supports the robustness of our results.

The frequency-domain tensor amplitude is modified by both the effective gravitational coupling, $G/G_N=(1+\kappa)^{-1}$, and the SPA factor $\dot f^{-1/2}$. Their combined contribution gives
\begin{equation}
C_\kappa
=\frac{1}{(1+\kappa)\sqrt{\mathcal{K}}}
=\frac{1}{\sqrt{(1+\kappa)(1+\kappa/6)}}.
\label{eq:KK_amplitude_factor}
\end{equation}
Consequently, the leading-order KK tensor waveform becomes
\begin{equation}
\widetilde h_{+/\times}^{\rm KK}(f)
=P_{+/\times}C_\kappa
\mathcal{A}_{\rm GR}(f)
e^{i[\Psi_{\rm GR}(f)+\beta_0u^{-5}]},
\label{eq:KK_waveform_0PN}
\end{equation}
where
\begin{equation}
\mathcal{A}_{\rm GR}(f)
=\sqrt{\frac{5}{24}}\pi^{-2/3}
\left(\frac{G_N\mathcal{M}_c}{c^3}\right)^{5/6}
\frac{c}{D_L}f^{-7/6},
\end{equation}
and
$P_+=(1+\cos^2\iota)/2$,
$P_\times=-i\cos\iota$.
Here $D_L$ is the luminosity distance and $\iota$ is the binary inclination. The corresponding leading-order ppE amplitude parameters are
$\alpha_0=C_\kappa-1$ and $a_0=0$.

We next incorporate the additional amplitude correction arising from perturbations of the compact internal volume. As derived in the preceding subsection, this contribution enters at relative 1PN order~\cite{Qiang:2016chy}. Retaining the volume-induced amplitude correction at this order, its relative correction is proportional to $8\kappa x$. Since the two contributions share the same stationary point, the leading SPA preserves this multiplicative amplitude dependence, giving
\begin{equation}
1+\delta A_{\rm vol}(f)
=1+8\kappa\eta^{-2/5}u^2.
\label{eq:KK_volume_ppe}
\end{equation}

Combining the leading-order KK waveform with the relative 1PN volume-induced amplitude correction, we obtain the effective waveform:
\begin{align}
\widetilde h_{+,\times}^{\rm KK}(f)
&=P_{+,\times}\mathcal{A}_{\rm GR}(f)C_\kappa
\left[1+8\kappa\eta^{-2/5}u^2\right]
\nonumber\\
&\quad\times
\exp\left\{i\left[
\Psi_{\rm GR}(f)+\beta_0u^{-5}
\right]\right\}.
\label{KK_full_effective}
\end{align}
Equivalently, writing
$\widetilde h^{\rm KK}
=\widetilde h^{\rm GR}(1+\delta A_{\rm eff})
e^{i\delta\Psi_{\rm eff}}$,
the combined ppE corrections are
\begin{align}
\delta A_{\rm eff}
&=\alpha_0+\alpha_2u^2,
\nonumber\\
\delta\Psi_{\rm eff}
&=\beta_0u^{-5},
\end{align}
with
\begin{align}
\alpha_0&=C_\kappa-1,
& a_0&=0,
\nonumber\\
\alpha_2&=8\kappa C_\kappa\eta^{-2/5},
& a_2&=2,
\nonumber\\
\beta_0&=\frac{15\kappa}{128(6+\kappa)},
& b_0&=-5.
\end{align}
All three coefficients vanish in the GR limit with $\kappa=0$.

The waveform constructed here incorporates the leading-order KK corrections and the 1PN amplitude contribution from internal volume perturbations. At Newtonian order, the phase correction is degenerate with a redefinition of the chirp mass, potentially limiting constraints on $\kappa$. A complete 1PN treatment of the conservative dynamics and GW energy flux, including the associated phase correction, may help break this degeneracy and will be pursued in future work.

\section{Multimessenger GW Events}
\label{multimessenger_events}

In this section, we introduce the multimessenger GW events considered in this work. We first summarize the key properties of GW170817, the first observed binary neutron star merger, and its electromagnetic counterpart~\cite{LIGOScientific:2017vwq}. We then discuss the astrophysical origins, electromagnetic detection strategies, and expected event rates of multimessenger MBHB mergers accessible to future space-based GW detectors~\cite{Mangiagli:2022niy}. Based on these considerations, we specify the source parameters for a representative multimessenger MBHB injection in the Taiji detector.

\begin{figure*}[t]
  \centering
  \begin{minipage}{0.48\textwidth}
    \centering
    \includegraphics[width=\textwidth]{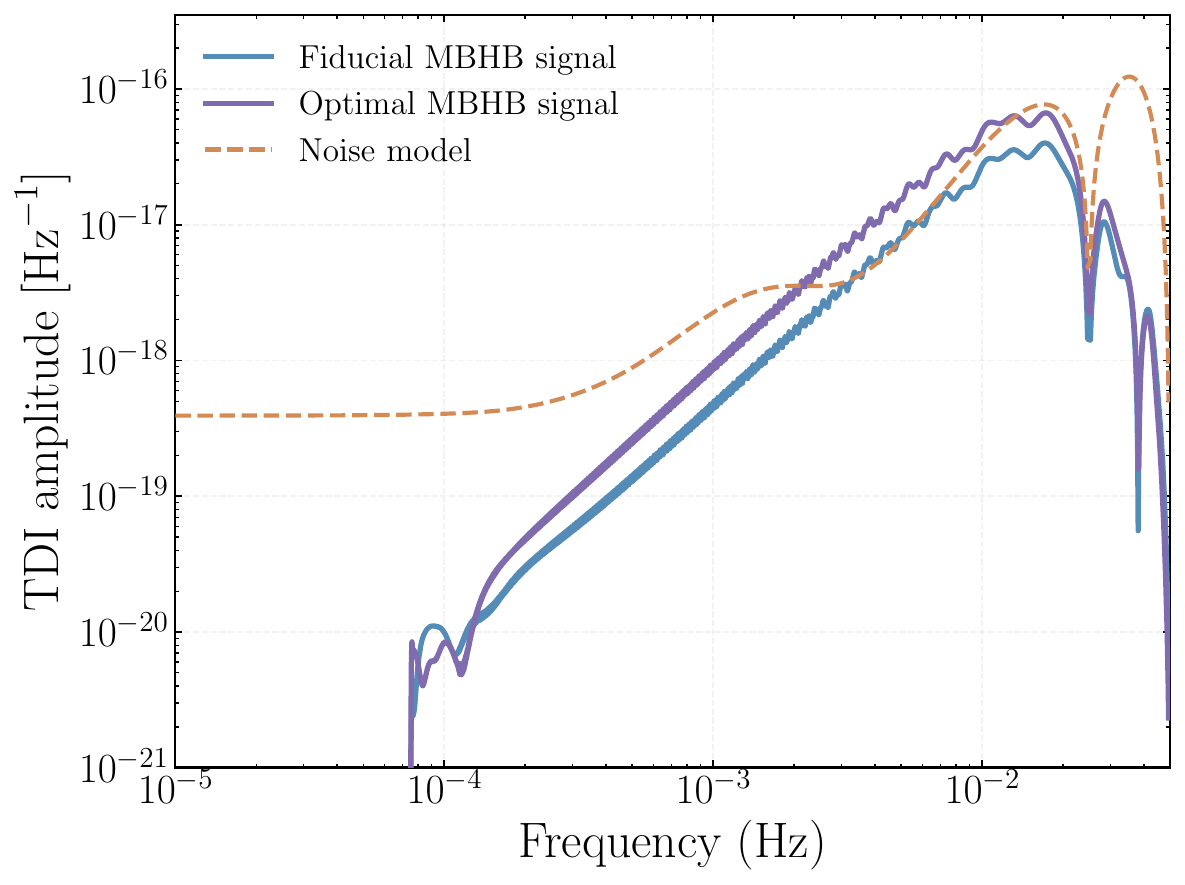}
  \end{minipage}
  \hfill
  \begin{minipage}{0.48\textwidth}
    \centering
    \includegraphics[width=\textwidth]{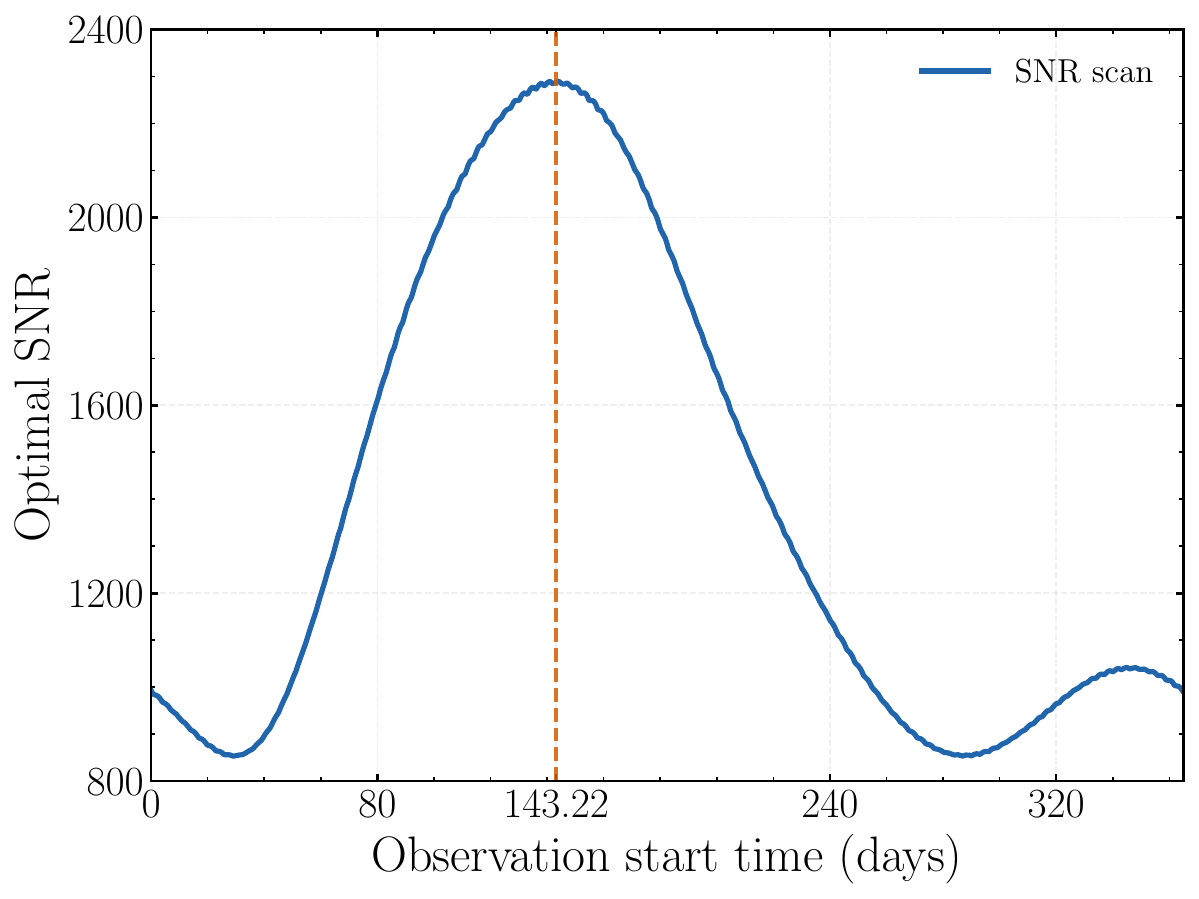}
  \end{minipage}
  \caption{{Injected MBHB signals and optimization of the Taiji observational configuration.}
  The left panel shows the frequency-domain GW signals for the fiducial and optimal MBHB injection configurations in the Taiji TDI channel. The solid blue curve represents the injected fiducial MBHB signal generated using the \textsc{IMRPhenomHM} waveform and propagated through the Taiji response, while the purple curve corresponds to the injected MBHB signal under the optimal observational configuration. The dashed orange curve indicates the adopted total noise level, including both instrumental and Galactic confusion noise.
  The right panel shows the event SNR as a function of the observation start time for a fixed one-year observation, with the merger occurring 350 days after the observation begins. The vertical dashed orange line marks the start time that maximizes the SNR.}
  \label{MBHB_SNR_optimization}
\end{figure*}

\subsection{Multimessenger observation of GW170817}
On August 17, 2017, Advanced LIGO and Advanced Virgo observed GW170817, the first GW signal from a BNS coalescence. The event was detected with a network signal-to-noise ratio of 32.4. The GW data localized the source to a 90\% credible sky area of $28~{\rm deg}^{2}$ and a three-dimensional volume of approximately $380~{\rm Mpc}^{3}$~\cite{LIGOScientific:2017vwq,LIGOScientific:2017zic,LIGOScientific:2017ync}. Approximately $1.7~{\rm s}$ after the merger, the associated short-duration gamma-ray burst GRB~170817A was detected by the Fermi Gamma-ray Burst Monitor~\cite{Goldstein:2017mmi}. Within 11 hours, the bright optical transient AT~2017gfo was identified in the host galaxy NGC~4993~\cite{Pan:2017jem}. Subsequent observations across the electromagnetic spectrum established its association with GW170817 and provided independent information about the source position, distance, and inclination~\cite{LIGOScientific:2017ync}. In particular, Hubble Space Telescope measurements of surface-brightness fluctuations in NGC~4993 yielded $D_{L}=40.7\pm1.4\pm1.9~{\rm Mpc}$, where the two uncertainties are statistical and systematic, respectively~\cite{Cantiello:2018ffy}. Dark Energy Camera observations in the $i$ and $z$ bands identified the optical counterpart and precisely determined its sky position~\cite{DES:2017kbs}. 

For GW170817, we use the publicly available cleaned strain data from the LIGO Hanford (H1) and Livingston (L1) detectors provided by GWOSC~\cite{GWOSC_GWTC}. We further fix the luminosity distance to $D_L=40.7~{\rm Mpc}$ and the right ascension and declination to $\alpha=3.45~{\rm rad}$ and $\delta=-0.41~{\rm rad}$, respectively, based on independent electromagnetic observations.

\subsection{Multimessenger Observation of injected MBHBs with Taiji}
\label{MBHB_event}

Several space-based GW observatories are planned for deployment in the 2030s, including LISA~\cite{LISA:2017pwj}, Taiji~\cite{Hu:2017mde,Du:2025xdq, Liang:2025izu,Liu:2026bbp}, and TianQin~\cite{TianQin:2015yph}. To assess the multimessenger prospects for MBHBs, Ref.~\cite{Mangiagli:2022niy} considered three representative black-hole seeding and merger-delay models. The Pop3 model assumes light seeds formed through the collapse of very massive, metal-poor Population III stars at high redshift, with merger delays included~\cite{Madau:2001sc,Volonteri:2002vz}. The Q3d model adopts heavy seeds formed through the direct collapse of protogalactic disks and includes merger delays~\cite{Begelman:2006db,Volonteri:2007ax}, while Q3nd follows the same heavy-seed prescription but neglects such delays~\cite{Begelman:2006db,Volonteri:2007ax}.

Using semianalytic galaxy-formation catalogs, Ref.~\cite{Mangiagli:2022niy} simulated MBHB populations over a four-year observing period. The optical, radio, and soft X-ray emission was then modeled using the gas content, black-hole masses and spins, and host-galaxy properties provided by these catalogs. The detectability of these signals was assessed for the Vera C. Rubin Observatory~\cite{LSST:2008ijt}, the Square Kilometre Array (SKA)~\cite{Dewdney:2009tmd}, and the Advanced Telescope for High Energy Astrophysics (Athena)~\cite{Nandra:2013jka}, with the source redshift determined either directly by Rubin or through follow-up observations with the Extremely Large Telescope~\cite{Padovani:2023dxc}.  In the maximizing scenario, which assumes isotropic radio-flare emission and neglects obscuration by gas and dust, the expected numbers of electromagnetic counterparts that can be sufficiently well localized by the space-based GW observatory and for which a redshift can be determined are $6.8$, $14.9$, and $20.9$ for the Pop3, Q3d, and Q3nd models, respectively~\cite{Mangiagli:2022niy}.

Following Ref.~\cite{Mangiagli:2022niy}, we adopt a representative set of parameters for MBHBs observable by Taiji~\cite{Du:2025xdq}. The fiducial source parameters are chosen as $\mathcal{M}_{c,z}=4\times10^{5}\,M_{\odot}$, $q=0.7$, $\chi_{1z}=\chi_{2z}=0.9$, $z=3$, $\iota=\pi/3$,  $\alpha=\pi/7$, $\delta=\pi/4$, $\psi=\pi/3$, and $\phi_{\rm ref}=\pi/5$.~\footnote{For convenience, we continue to use $(\alpha,\delta)$ to denote the source sky position for space-based gravitational-wave detectors.
Unlike in the ground-based case, these angles refer to the ecliptic longitude and latitude, respectively.}
Here, $\mathcal{M}_{c,z}$ denotes the detector-frame chirp mass, $q=m_2/m_1\leq1$ the mass ratio, $\chi_{1z}$ and $\chi_{2z}$ the dimensionless spin components aligned with the orbital angular momentum, and $z$ denotes the source redshift, with $D_L$ determined from $z$ assuming a $\Lambda$CDM cosmology.  The angles $\iota$, $(\alpha,\delta)$, and $\psi$ denote the binary inclination, the ecliptic longitude and latitude, and the polarization angle, respectively.  For this multimessenger MBHB event, independent electromagnetic observations can provide additional constraints on the luminosity distance $D_L$ and the source sky position $(\alpha,\delta)$.

As shown in the left panel of Fig.~\ref{MBHB_SNR_optimization}, to obtain a more realistic injected data stream, we additionally
generate the instrumental noise expected for Taiji using the
\textsc{Triangle} framework~\cite{Du:2025xdq}. We also include the
confusion foreground produced by unresolved Galactic compact binaries,
using the Taiji confusion-noise model of Ref.~\cite{Liu:2023qap}. The dashed orange curve in the left panel shows the total noise level. The simulated data in each TDI channel are therefore given by
\begin{equation}
\tilde d_I(f) = \tilde h_I^{\rm GR}(f) +\tilde n_I^{\rm inst}(f) +\tilde n_I^{\rm conf}(f),
\end{equation}
where $\tilde n_I^{\rm inst}$ and $\tilde n_I^{\rm conf}$ denote the instrumental and Galactic-confusion noise contributions, respectively.  Throughout this work, we use only the $A$ and $E$ TDI channels~\cite{Marsat:2018oam, Marsat:2020rtl, Du:2025xdq}, for which the injected signal has an SNR of 981.

We simulate one year of MBHB observations with a sampling interval of 10 s and construct the frequency-domain data using the corresponding fast Fourier transform (FFT). To improve computational efficiency, we exclude frequency bins below $10^{-5}~\mathrm{Hz}$, which lie outside the sensitive band of Taiji. We assume that the MBHB merger occurs 350 days after the start of the observation. 
For the Taiji constellation, we adopt the analytical equal-arm orbital configuration~\cite{Rubbo:2003ap, Wu:2019thj} implemented in \textsc{Triangle-BBH}, a simulation framework for modeling MBHB GW signals and the Taiji detector response~\cite{ Du:2025xdq}.
Owing to Taiji's one-year orbital period, varying the observation start time changes the detector's orbital phase and its time-dependent antenna response to the source, thereby modulating the event SNR, as shown in the right panel of Fig.~\ref{MBHB_SNR_optimization}. The vertical orange dashed line marks the optimal observational configuration that maximizes the SNR. Accordingly, we consider two injection configurations, with the fiducial case starting at 0 day and yielding an SNR of 981, while the optimal case starts at 143.22 days and achieves an SNR of 2290. The corresponding frequency-domain signals are shown by the solid blue and purple curves, respectively, in the left panel.

\section{Implementation of Extra-Dimensional Waveform Templates and Bayesian Inference}
\label{extra_dimensional_templates}

In this section, we first describe the implementation of extra-dimensional corrections to the \textsc{TaylorF2} waveform for GW170817 observed by ground-based detectors. For MBHB observations with Taiji, the exceptionally high SNR motivates the use of the \textsc{IMRPhenomHM} waveform, which incorporates the full inspiral–merger–ringdown evolution. We extend the extra-dimensional corrections from the dominant $(2,2)$ mode to higher-order $(\ell,m)$ modes and introduce Planck-taper windows for both the amplitude and phase corrections to confine their effects to the inspiral regime. Finally, we present the Bayesian inference framework used to constrain extra dimensions from GW observations with these modified waveform templates.

\subsection{Extra-Dimensional Waveform Implementation for GW170817}
\label{GW170817_waveform}

The GW170817 event shows no significant evidence for spin precession, and previous analyses have found that precessing and nonprecessing waveform models yield consistent posterior distributions for this event~\cite{LIGOScientific:2018hze}. We therefore model GW170817 as a nonprecessing compact binary, allowing the component spins to be aligned or antialigned with the orbital angular momentum. In particular, we adopt \textsc{TaylorF2} as the baseline waveform model, retaining only the dominant $(2,2)$ mode and restricting the analysis to the inspiral phase. We can therefore directly employ the extra-dimensional waveform template given in Eq.~\ref{KK_full_effective}. The response of a ground-based GW detector is expressed as
\begin{equation}
\tilde{h}^{\rm KK}(f)=F_{+}(\alpha,\delta,\psi)\tilde{h}^{\rm KK}_{+}(f)+F_{\times}(\alpha,\delta,\psi)\tilde{h}^{\rm KK}_{\times}(f),
\end{equation}
where $F_{+}$ and $F_{\times}$ are the detector antenna pattern functions. Both the modified waveform and the detector response are implemented using \textsc{PyCBC}~\cite{Biwer:2018osg}.

\subsection{Extra-Dimensional Waveform Implementation for Taiji MBHBs}

The analysis of space-based GW data differs in several important respects from that of ground-based interferometers as mentioned in the previous subsection.
In particular, MBHBs can remain in band from months to years, allowing the orbital motion and evolving orientation of the detector constellation to imprint characteristic amplitude and phase modulations, as well as propagation delays, on the observed signal.
These effects can be efficiently incorporated in the frequency domain through frequency-dependent transfer functions~\cite{Marsat:2018oam}. Moreover, the fundamental observables of space-based interferometers are time-delay interferometry (TDI) combinations~\cite{Tinto:2020fcc, Wang:2017aqq, Shaddock:2003dj, Tinto:2022zmf}, rather than the GW strain measured by an individual interferometer.

In this work, we construct a simulated multimessenger MBHB observation
using \textsc{Triangle-BBH}~\cite{Du:2025xdq}. We first generate the frequency-domain gravitational waveform using the aligned-spin \textsc{IMRPhenomHM} model, including the $(\ell,m)=(2,2),(2,1),(3,3),(3,2),(4,4)$, and $(4,3)$ modes. Since the ppE corrections derived in Sec.~\ref{ppE_waveform} apply only to the inspiral and the dominant $(2,2)$ mode, two extensions are required to incorporate them into the full inspiral-merger-ringdown (IMR) waveform.

First, following Mezzasoma and Yunes~\cite{Mezzasoma:2022pjb}, who extended the ppE framework to include higher harmonics with $\ell=2,3,4$, we generalize the phase correction to each $(\ell,m)$ mode. Motivated by the common orbital evolution of the harmonics, we also adopt a phenomenological prescription for their fractional amplitude corrections. The resulting mappings are~\cite{Mezzasoma:2022pjb, Mehta:2022pcn, Xie:2024xex, Piarulli:2025rvr}
\begin{align}
\delta\Psi_{\ell m}(f)
&=\frac{m}{2}\delta\Psi_{22}\left(\frac{2f}{m}\right),
\nonumber\\
\delta A_{\ell m}(f)
&=\delta A_{22}\left(\frac{2f}{m}\right).
\label{eq:KK_HM_mapping}
\end{align}
Substituting the ppE corrections derived in Sec.~\ref{ppE_waveform}, we obtain
\begin{align}
\delta\Psi_{\ell m}(f)
&=\beta_0\left(\frac{m}{2}\right)^{8/3}u^{-5},
\nonumber\\
\delta A_{\ell m}(f)
&=\alpha_0+\alpha_2
\left(\frac{2}{m}\right)^{2/3}u^2.
\label{KK_HM_corrections}
\end{align}

For space-based GW detectors such as LISA, Taiji and TianQin, the phase correction also modifies the time-frequency correspondence entering the time-dependent detector response. Within the SPA, this relation becomes~\cite{Piarulli:2025rvr}
\begin{align}
\delta t_{\ell m}(f)
&=\frac{1}{2\pi}
\frac{d \delta \Psi_{\ell m}(f)}{df}.
\end{align}

Second, since the ppE corrections derived above are valid only during the inspiral stage, we smoothly taper off the extra-dimensional modifications before merger, following the frequency-domain prescription of Piarulli et al.~\cite{Piarulli:2025rvr}. For the dominant $(2,2)$ mode, we adopt $f_{\rm taper}=0.35f_{\rm peak}^{22}$, where $f_{\rm peak}^{22}$ denotes the peak frequency of the $(2,2)$ mode, and define the transition interval $[f_a,f_b]$ with $f_a=0.9f_{\rm taper}$ and $f_b=1.1f_{\rm taper}$. We employ the descending Planck-taper window~\cite{McKechan:2010kp},
\begin{equation}
W(f)=
\begin{cases}
1, & f\leq f_a,\\[3pt]
\displaystyle
\left[1+e^{Q(f)}\right]^{-1},
& f_a<f<f_b,\\[3pt]
0, & f\geq f_b,
\end{cases}
\end{equation}
with 
\begin{equation}
Q(f)=
\frac{f_a-f_b}{f-f_a}
+\frac{f_a-f_b}{f-f_b}.
\end{equation}
This choice smoothly suppresses the inspiral corrections while retaining the GR merger-ringdown waveform. The tapered phase is reconstructed as
\begin{align}
\delta\Psi_{22}^{\rm tap}(f)
&=\int_{f_b}^{f}df'\int_{f_b}^{f'}df''\,
W(f'')\frac{d^2\delta\Psi_{22}^{\rm ppE}(f'')}{df''^2}
\nonumber\\
&=\int_{f_b}^{f}(f-f')W(f')
\frac{d^2\delta\Psi_{22}^{\rm ppE}(f')}{df'^2}\,df'.
\label{eq:KK_phase_taper}
\end{align}
The integration boundary conditions ensure $\delta\Psi_{22}^{\rm tap}(f_b)=0$ and $d\delta\Psi_{22}^{\rm tap}/df|_{f_b}=0$, allowing a smooth transition to the GR merger-ringdown phase. Below $f_a$, the correction takes the form
\begin{equation}
\delta\Psi_{22}^{\rm tap}(f)
=\delta\Psi_{22}^{\rm ppE}(f)
+A+B(f-f_a),
\qquad f\leq f_a,
\label{eq:KK_phase_low_frequency}
\end{equation}
where $A$ and $B$ are fixed by requiring continuity of the phase and its first derivative at $f_a$. These terms account for the phase and time alignment introduced by the integration procedure.

For the amplitude, we apply the window directly to the relative ppE correction, rather than to the GR amplitude, following the phenomenological tapering approach used in modified-gravity waveform analyses~\cite{LISACosmologyWorkingGroup:2026zah}. The resulting amplitude correction is
\begin{equation}
\delta A_{22}^{\rm tap}(f)
=W(f)\delta A_{22}^{\rm ppE}(f)
=W(f)\left(\alpha_0+\alpha_2u^2\right).
\label{eq:KK_amplitude_taper}
\end{equation}
Thus, the original inspiral correction is preserved below $f_a$, whereas the GR amplitude is recovered above $f_b$.

Finally, the tapered corrections are extended to higher harmonics using the frequency mappings described above~\cite{Mezzasoma:2022pjb,Piarulli:2025rvr},
\begin{align}
\delta\Psi_{\ell m}^{\rm tap}(f)
&=\frac{m}{2}\delta\Psi_{22}^{\rm tap}
\left(\frac{2f}{m}\right),
\nonumber\\
\delta A_{\ell m}^{\rm tap}(f)
&=\delta A_{22}^{\rm tap}
\left(\frac{2f}{m}\right).
\label{eq:KK_HM_taper}
\end{align}
The resulting effective KK waveform is
\begin{equation}
\widetilde h_{\ell m}^{\rm KK}(f)
=\widetilde h_{\ell m}^{\rm GR}(f)
\left[1+\delta A_{\ell m}^{\rm tap}(f)\right]
e^{i\delta\Psi_{\ell m}^{\rm tap}(f)}.
\label{eq:KK_IMR_waveform}
\end{equation}

For a space-based GW detector like Taiji, the
response must account for the motion of the spacecraft constellation
throughout the signal.  Following the frequency-domain response formalism
of Ref.~\cite{Marsat:2018oam},  the response in a TDI channel $I$ can be
schematically written as~\cite{Marsat:2020rtl},
\begin{equation}
\tilde h_I^{\rm KK}(f) = \sum_{\ell,m} \mathcal{T}^{I}_{\ell m}  \left(f;\alpha,\delta,\psi,\iota\right) \tilde h_{\ell m}^{\rm KK}(f),
\end{equation}
where $\tilde h_{\ell m}$ are the individual frequency-domain
\textsc{IMRPhenomHM} modes in the KK theory, and
$\mathcal{T}^{I}_{\ell m}$ represents the frequency-dependent detector
transfer function, including the orbital modulation, propagation delays,
and the TDI combination. Using \textsc{Triangle-BBH}, we propagate the
source waveform through the time-dependent Taiji constellation and
construct the second-generation TDI observables~\cite{Du:2025xdq}. We thus obtain an effective multimode KK GW template covering the full IMR regime for space-based GW detectors. To assess Taiji's ability to probe extra dimensions, we first generate injected GW data assuming GR and then perform parameter inference using the KK waveform template.

\subsection{Bayesian Parameter Estimation}
In this work, we employ Bayesian inference to constrain both the standard GR parameters and the additional extra-dimensional parameters from GW observations~\cite{Finn:1992wt}. Given the observed data $\tilde{d}$ in frequency domain and a hypothesis $\mathcal{H}$, Bayes' theorem gives the posterior distribution of the parameter vector $\vartheta$ as
\begin{equation}
p(\vartheta|\tilde{d},\mathcal{H})=\frac{p(\tilde{d}|\vartheta,\mathcal{H})\mathrm{~}p(\vartheta|\mathcal{H})}{p(\tilde{d}|\mathcal{H})},
\end{equation}
where $p(\tilde{d}|\vartheta,\mathcal{H})$ is the likelihood, $p(\vartheta|\mathcal{H})$ is the prior distribution, and $p(\tilde{d}|\mathcal{H})$ is the Bayesian evidence. Assuming stationary Gaussian noise that is statistically independent across ground-based detectors, the joint log-likelihood takes the form
\begin{equation}
\log p(\tilde{d}|\vartheta,\mathcal{H})=\log\bar{\alpha}-\frac{1}{2}\sum_i\left\langle \tilde{d}_i-\tilde{h}^{\rm KK}_i|\tilde{d}_i-\tilde{h}^{\rm KK}_i\right\rangle,
\end{equation} 
where $\bar{\alpha}$ is a normalization constant, $i$ labels the detectors, and $\tilde{h}^{\rm KK}_i$ denotes the extra dimensional waveform in KK gravity projected onto the $i$th detector. The noise-weighted inner product is defined as
\begin{equation}
    \langle a(t)|b(t)\rangle=2\int_{f_{\mathrm{low}}}^{f_{\mathrm{high}}}\frac{\tilde{a}^*(f)\tilde{b}(f)+\tilde{a}(f)\tilde{b}^*(f)}{S_n(f)}df,
\end{equation}
where $f_{\mathrm{low}}$ and $f_{\mathrm{high}}$ are the low- and high-frequency cutoffs of the data, respectively.

For the Taiji MBHB analysis, we employ the \textsc{Triangle-BBH} package to compute the frequency-domain second-generation TDI response~\cite{Du:2025xdq,Marsat:2020rtl}.
The response incorporates the orbital motion of the Taiji constellation and the finite-arm transfer functions. The resulting TDI Michelson observables $X$, $Y$, and $Z$ are combined into the $A$ and $E$ channels. In this work, we restrict our analysis to the $A$ and $E$ channels and define the data and waveform template vectors as
$\tilde{\boldsymbol d} =(\tilde d_A,\tilde d_E)^{\rm T}$ and $\tilde{\boldsymbol h}^{\rm KK} =(\tilde h_A^{\rm KK}, \tilde h_E^{\rm KK})^{\rm T}$.
The corresponding joint log-likelihood is
\begin{equation}
\ln p(\tilde{\boldsymbol d}|\vartheta,\mathcal H) 
=\ln\bar{\alpha}_{\rm TDI} -\frac{1}{2}\left\langle\tilde{\boldsymbol d}
-\tilde{\boldsymbol h}^{\rm KK}\middle|\tilde{\boldsymbol d}
-\tilde{\boldsymbol h}^{\rm KK}\right\rangle_{\rm TDI},
\end{equation}
where the multi-channel noise-weighted inner product is defined as
\begin{equation}
\begin{aligned}
\langle\boldsymbol a|\boldsymbol b\rangle_{\rm TDI}=4\,\mathrm{Re}
\int_{f_{\rm low}}^{f_{\rm high}}\tilde{\boldsymbol a}^{\dagger}(f)\times\boldsymbol S_n^{-1}(f)\tilde{\boldsymbol b}(f)\,df.
\end{aligned}
\end{equation}
Here $\boldsymbol S_n(f)$ denotes the one-sided noise cross-spectral density matrix of the $A$ and $E$ channels. We assume that the noise in these two channels is uncorrelated and that their noise PSDs are approximately equal, $S_A(f)=S_E(f)\simeq S_{\rm inst}(f)+S_{\rm conf}(f)$, where $S_{\rm inst}(f)$ and $S_{\rm conf}(f)$ represent the instrumental and Galactic confusion-noise contributions, respectively.

In this work, we model both GW170817 and the MBHB system as quasicircular, spin-aligned compact binaries. The extra-dimensional waveform in KK gravity is therefore characterized by the following 12 parameters,
\begin{equation}
\vartheta_{\rm KK}=\{\mathcal{M}_c,q,\chi_{1z},\chi_{2z},D_L,\iota, \alpha,\delta,\psi,t_c,\phi_c, \kappa\},
\end{equation}
where $\mathcal{M}_c$ and $q$ are chirp mass and the mass ratio,
$\chi_{1z}$ and $\chi_{2z}$ are the dimensionless spin components along the orbital angular momentum, $D_L$ is the luminosity distance of the source, ($\alpha$, $\delta$) specify the source sky position, $\iota$ and $\psi$ are the inclination angle and the polarization angle respectively, $\phi_c$ and $t_c$ are the coalescence phase and time, respectively, and $\kappa$ is the additional parameter characterizing the extra-dimensional effects. In the subsequent parameter estimation, we assume the identification of an electromagnetic counterpart and fix the luminosity distance and sky position $(D_L, \alpha,\delta)$, leaving nine free parameters to be inferred.

For the GW170817 analysis, we adopt uniform priors $\mathcal{M}_c\in[1.15,1.23]~M_\odot$, $q\in[1,2]$, $\chi_{1z},\chi_{2z}\in[-0.05,0.05]$, $\psi,\phi_c\in[0,2\pi]$, and $\delta t_c\in[-0.1,0.1]~{\rm s}$.
For the inclination angle $\iota$, we adopt a sine prior $p(\iota)\propto\sin\iota$ with $\iota\in[0,\pi]$.
For the extra-dimensional parameter, we adopt a uniform prior $\kappa\in(0,1)$.  We perform Bayesian inference using the \textsc{PyCBC} framework~\cite{Biwer:2018osg}, with \textsc{Bilby} employed for posterior sampling~\cite{Ashton:2018jfp,Romero-Shaw:2020owr}. 

For the parameter estimation of the simulated multimessenger MBHB, we first use the Fisher information matrix (FIM)~\cite{Vallisneri:2007ev} implemented in \textsc{Triangle-BBH}~\cite{Du:2025xdq} to estimate the characteristic $1\sigma_{\rm FIM}$ uncertainties around the injected parameters. The Fisher matrix is employed only to construct computationally efficient priors and is not used to derive the final parameter constraints. Specifically, for the detector-frame chirp mass $\mathcal{M}_c$, mass ratio $q$~
\footnote{Following the conventions adopted in \textsc{PyCBC} and \textsc{Triangle-BBH}, we use different definitions of the mass ratio $q$ for GW170817 and MBHB systems. Assuming $m_1>m_2$, we define $q=m_1/m_2$ for GW170817 and $q=m_2/m_1$ for MBHBs, respectively.}
, aligned spin components $\chi_{1z}$ and $\chi_{2z}$, and reference time $t_{\rm ref}$, we adopt uniform priors centered on the injected values with boundaries extending to $\pm10\sigma_{\rm FIM}$, additionally restricted by the corresponding physical ranges. For the remaining free parameters, we adopt standard physical priors such as $\phi_{\rm ref}\in[0,2\pi)$, an isotropic orientation prior for $\iota$, $\psi\in[0,\pi)$, and a uniform prior $\kappa\in(0,1)$ for the extra-dimensional parameter. 

To reduce the computational cost associated with repeatedly evaluating the year-long Taiji signal and its second-generation TDI response, we employ the heterodyned-likelihood method~\cite{Cornish:2021lje}. Taking the injected GR waveform as the reference waveform, the likelihood is compressed onto $256$ heterodyne frequency nodes, while retaining the information contained in the original data. This treatment substantially accelerates likelihood evaluations in the neighborhood of the high-likelihood region. The posterior distribution is then sampled using the normalizing-flow-based nested sampler \textsc{Nessai}~\cite{Williams:2021qyt} through the \textsc{Bilby} framework~\cite{Ashton:2018jfp}.

\begin{figure}[ht] 
\centering 
\includegraphics[width=0.45\textwidth]{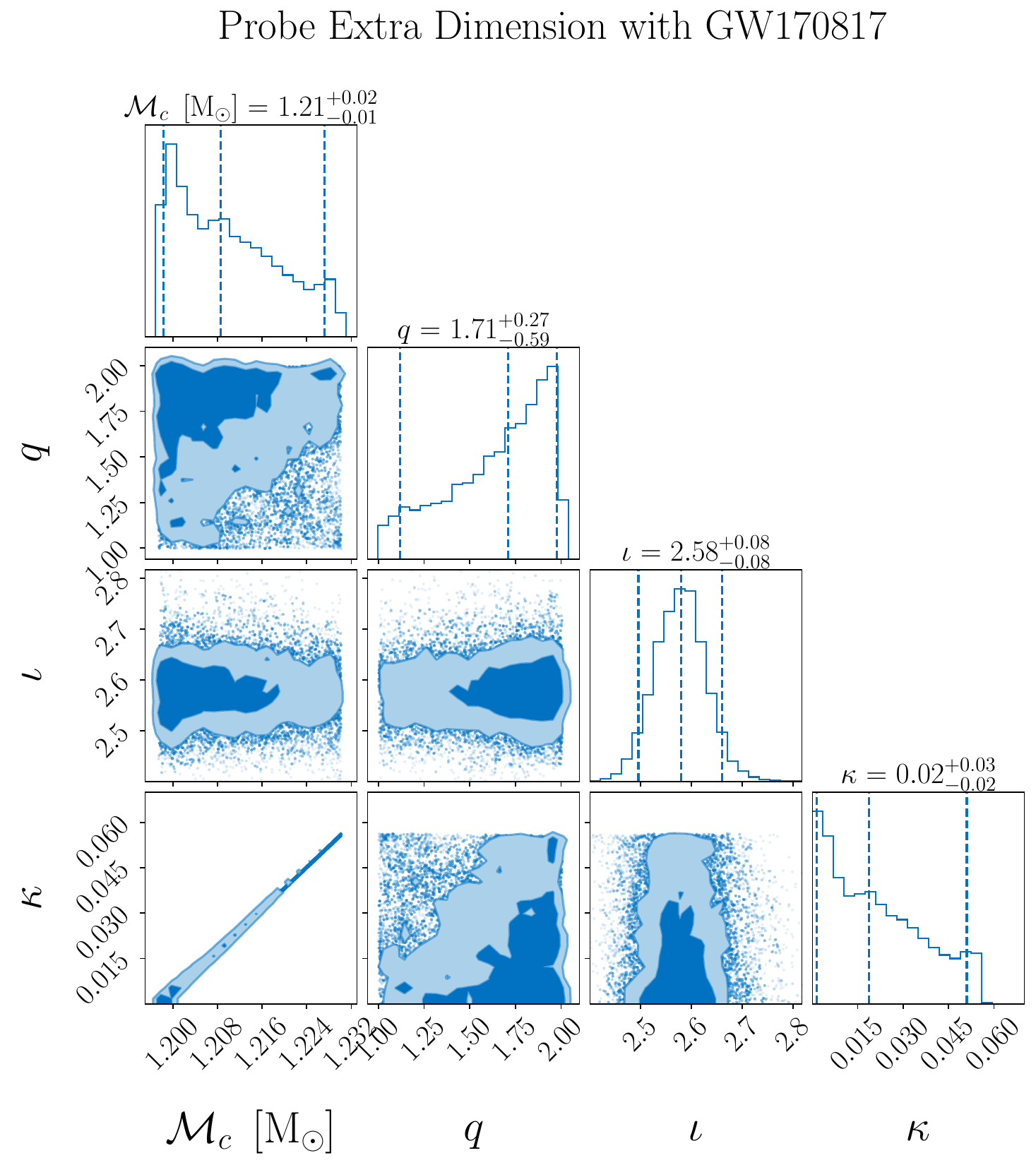} 
\caption{Posterior distributions for selected GR parameters and the extra-dimensional parameter $\kappa$ inferred from GW170817. The two-dimensional contours show the 50\% and 90\% credible regions, while in the one-dimensional marginalized posteriors the outer dashed vertical lines denote the 90\% credible interval and the central dashed vertical line marks the median.} 
\label{170817_results} 
\end{figure}

\section{Results and Discussion}
\label{results}

\begin{figure*}[t]
  \centering
  \begin{minipage}{0.48\textwidth}
    \centering
    \includegraphics[width=\textwidth]{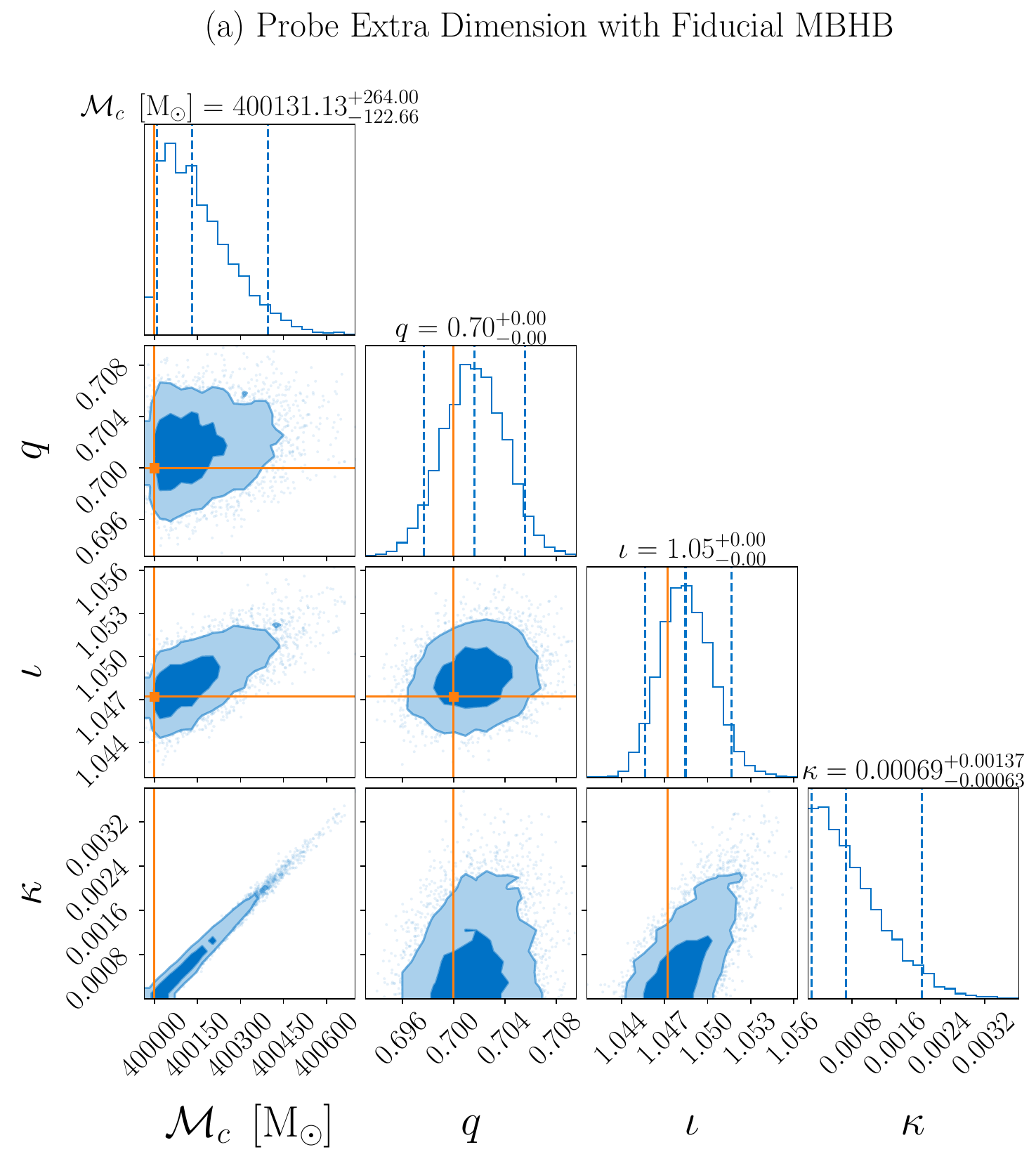}
  \end{minipage}
  \hfill
  \begin{minipage}{0.48\textwidth}
    \centering
    \includegraphics[width=\textwidth]{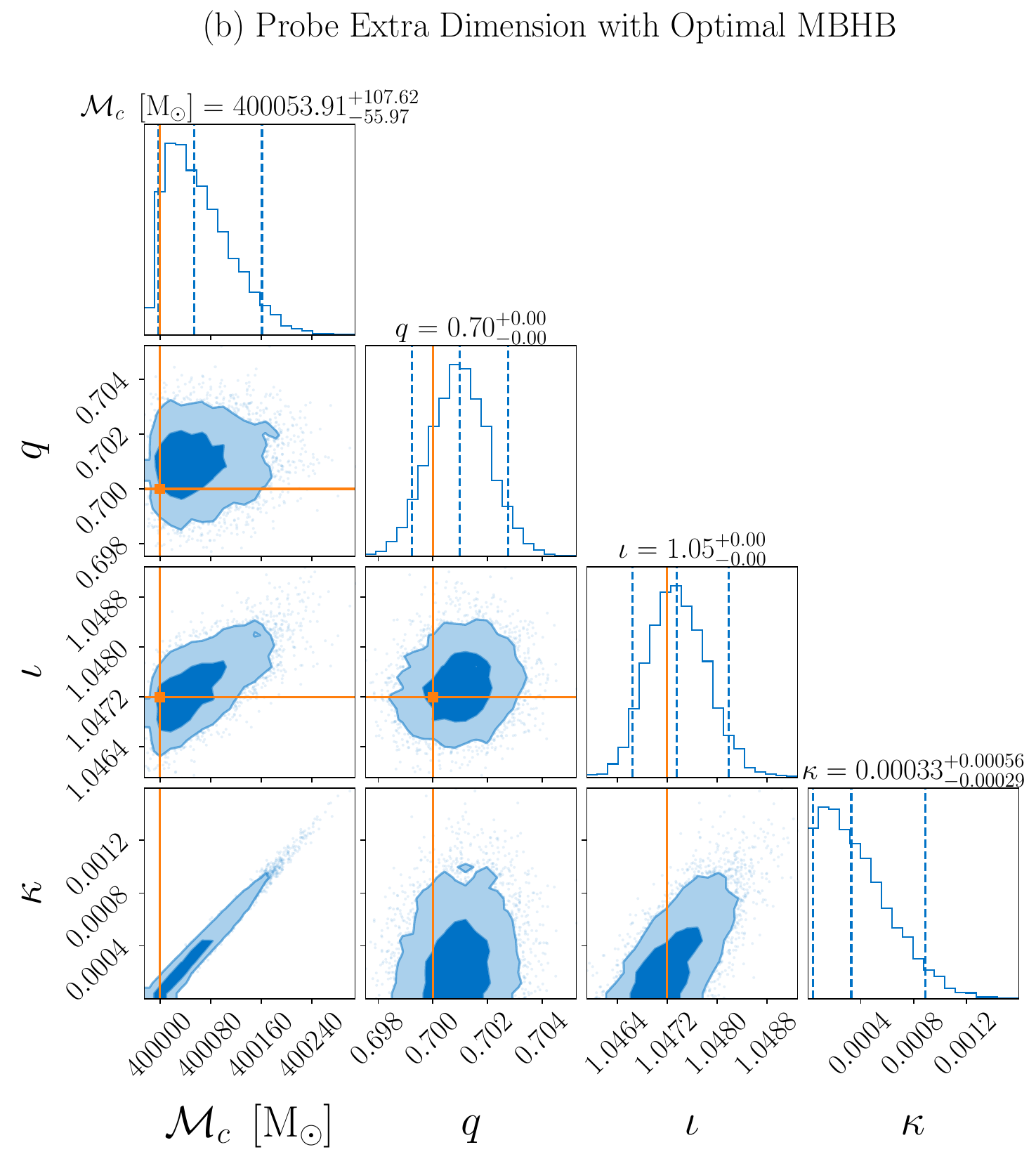}
  \end{minipage}
  \caption{Posterior distributions of selected binary parameters and the extra-dimensional parameter $\kappa$ for injected multimessenger MBHB observations with Taiji. The left and right panels correspond to the fiducial and optimal configurations, with SNRs of 981 and 2290, respectively. The orange solid lines indicate the injected parameter values.}
  \label{MBHB_results}
\end{figure*}

We first constrain the extra-dimensional effects using the observational data of GW170817, with the results presented in Fig.~\ref{170817_results}. The figure shows the posterior distributions of key GR parameters, including the chirp mass, mass ratio, and inclination, together with the extra-dimensional parameter $\kappa$. The two-dimensional contours enclose the 50\% and 90\% credible regions, while the vertical dashed lines in the one-dimensional marginalized distributions mark the median and the 90\% credible interval.

For the GR parameters, we obtain $\mathcal{M}_c=1.21^{+0.02}_{-0.01}~M_{\odot}$, $q=1.71^{+0.27}_{-0.59}$, and $\iota=2.58^{+0.08}_{-0.08}$. Although these values exhibit slight deviations from previous GR-based analyses~\cite{LIGOScientific:2017vwq,LIGOScientific:2017zic,LIGOScientific:2017ync,Nitz:2021zwj}, their credible intervals remain broadly consistent, providing a consistency check of our inference framework. For the extra-dimensional parameter, we obtain $\kappa=0.02^{+0.03}_{-0.02}$, corresponding to a 95\% credible upper bound of $n<0.11$ on the phenomenological dimensional parameter. This provides the direct constraint on $n$ in KK theory derived from real GW observations. The result is consistent with a four-dimensional spacetime, as assumed in standard GR, with no significant evidence for extra-dimensional effects.

Complementary GW-based tests of extra dimensions have also been explored within different theoretical frameworks. These include comparing GW and electromagnetic luminosity distances to search for gravitational leakage into extra dimensions~\cite{Pardo:2018ipy}, using the arrival-time difference between GWs and their electromagnetic counterparts to constrain higher-dimensional propagation effects~\cite{Lin:2023zna}, and searching for additional GW polarization modes arising from extra spatial dimensions~\cite{Wang:2026sir}. Despite their distinct theoretical motivations, these studies have found no compelling evidence for extra-dimensional effects, with their results remaining consistent with an effectively four-dimensional spacetime, as described by GR. 

Meanwhile, the extra-dimensional effects considered here within the KK framework can also be tested through Solar System observations of the parametrized post-Newtonian (PPN) parameter $\gamma_{\rm PPN}$. The static weak-field solution predicts $\gamma_{\rm PPN}=(1-\kappa)/(1+\kappa)=1/(n+1)$. Since Solar System observations require $\gamma_{\rm PPN}$ to be extremely close to unity, models with $n>0$ are strongly constrained and effectively ruled out except in the limit $n\to0$ under these assumptions~\cite{Eingorn:2010xp,Chopovsky:2013dg,Du:2020rlx}. While Solar System PPN tests impose stringent constraints in the static weak-field regime, GW observations provide a complementary probe of gravitational dynamics in compact binaries. By incorporating perturbations of the compact internal geometry into the generation of four-dimensional GWs, our analysis provides a direct phenomenological test of extra-dimensional effects in the strong-field, dynamical regime of compact-binary inspirals.

Future space-based GW detectors are expected to observe MBHB mergers with SNRs reaching several hundred or even several thousand. To investigate their potential for probing extra-dimensional effects in the strong-field, dynamical regime, we perform Bayesian inference on injected multimessenger MBHB observations with Taiji. For a given MBHB source, Taiji's orbital configuration modulates its time-dependent response and hence the accumulated event SNR. We therefore consider two injection scenarios with identical source parameters but different observation start times.

In the fiducial case, the observation begins at day 0 and lasts for one year, with the MBHB merger occurring on day 350, yielding an SNR of 981. The corresponding posterior distributions are shown in the left panel of Fig.~\ref{MBHB_results}. The injected values of the key binary parameters, including the chirp mass, mass ratio, and inclination, are well recovered. For the extra-dimensional parameter, we obtain $\kappa=0.00069^{+0.00137}_{-0.00063}$, corresponding to a 95\% credible upper bound of $n<0.00413$. This improves the constraint obtained from GW170817 by a factor of approximately 27, demonstrating the potential of high-SNR MBHB observations to probe extra-dimensional effects.

We further consider an optimal injection scenario by varying the observation start time to maximize the event SNR, as illustrated in Fig.~\ref{MBHB_SNR_optimization}. Keeping the one-year observation duration and the merger time fixed at 350 days after the observation begins, we find that an observation starting at day 143.22 yields the maximum SNR of 2290 for the same MBHB source. The corresponding posterior distributions, shown in the right panel of Fig.~\ref{MBHB_results}, again recover the injected binary parameters accurately. The extra-dimensional parameter is constrained to $\kappa=0.00033^{+0.00056}_{-0.00029}$, corresponding to a 95\% credible upper bound of $n<0.00178$. This represents a further two orders of magnitude improvement over the GW170817 results, highlighting the impact of Taiji's orbital configuration on its sensitivity to extra-dimensional effects.

Several factors are expected to contribute to the stronger constraints obtained from the Taiji injection. First, the one-year Taiji observations of the MBHB yield exceptionally high SNRs of 981 and 2290 for the fiducial and optimal cases, respectively, compared with a network SNR of approximately 32.4 for GW170817, enabling substantially more precise parameter estimation. Second, for the Taiji analysis we employ the \textsc{IMRPhenomHM} waveform, which includes the dominant $(\ell,m)=(2,2)$ mode together with the $(2,1)$, $(3,3)$, $(3,2)$, $(4,4)$, and $(4,3)$ higher-order modes, rather than the dominant-mode \textsc{TaylorF2} waveform adopted for GW170817. The distinct dependence of different $(\ell,m)$ modes on the inclination and mass ratio helps break degeneracies among these parameters in the waveform amplitude, thus providing additional information to constrain the extra-dimensional parameter $\kappa$. Moreover, the extra-dimensional amplitude and phase corrections acquire distinct mode-dependent factors, as shown in Eq.~\ref{KK_HM_corrections}, allowing different harmonics to probe the waveform modifications with different weights. Taken together, the high SNR and the additional information encoded in higher-order modes account for the substantially stronger constraint achievable with the Taiji MBHB observation, highlighting the unique potential of space-based GW detectors for precision tests of gravity.

It is worth noting that the leading-order amplitude and phase corrections, together with the 1PN amplitude correction derived within ppE framework, introduce strong correlations among $\kappa$, $D_L$, and the chirp mass $\mathcal{M}_c$ in parameter estimation. For the multimessenger events considered here, independent electromagnetic measurements of $D_L$ help break the degeneracy between $\kappa$ and $D_L$. However, a significant degeneracy between $\kappa$ and $\mathcal{M}_c$ remains, as illustrated in Figs.~\ref{170817_results} and \ref{MBHB_results}. Including higher-order extra-dimensional corrections, particularly the next-to-leading (1PN) phase correction, could introduce additional frequency dependence that helps disentangle $\kappa$ from $\mathcal{M}_c$. Such an extension could enable more robust joint constraints on extra-dimensional effects and the intrinsic binary parameters from multimessenger GW observations. We will investigate these next leading order corrections in future work.

In summary, we investigate the effects of compact extra dimensions on GW generation in the effective four-dimensional spacetime within the KK framework. We derive the extra-dimensional corrections to the conservative binary dynamics and GW energy flux and formulate the corresponding waveform modifications within the ppE framework. Combining these results with the previously derived 1PN amplitude correction arising from perturbations of the internal volume and the resulting modification of the effective four-dimensional stress-energy tensor, we construct extra-dimensional GW waveform templates for compact binaries. Using the observed strain data of GW170817 together with independent electromagnetic observations, our Bayesian analysis yields a 95\% credible upper bound of $n<0.11$ on the phenomenological dimensional parameter, providing a direct GW-based test of extra-dimensional effects in the strong-field, dynamical regime of compact-binary inspirals. We further investigate the prospects of future space-based GW detectors using injected multimessenger MBHB observations with Taiji. Our results show that high-SNR MBHB observations could improve the constraint by nearly two orders of magnitude, reaching $n<0.00178$. These findings highlight the potential of Taiji to probe extra-dimensional effects and test fundamental physics in the strong-field regime.

\emph{\textbf{Acknowledgements}.---}
This work is supported by the National Key R\&D Program of China under Grants No. 2025YFE0217300. 
This work is also supported by the High-performance Computing Platform of Peking University. 
This research has made use of data or software obtained from the Gravitational Wave Open Science Center (gwosc.org), a service of the LIGO Scientific Collaboration, the Virgo Collaboration, and KAGRA. This material is based upon work supported by NSF's LIGO Laboratory which is a major facility fully funded by the National Science Foundation, as well as the Science and Technology Facilities Council (STFC) of the United Kingdom, the Max-Planck-Society (MPS), and the State of Niedersachsen/Germany for support of the construction of Advanced LIGO and construction and operation of the GEO600 detector. Additional support for Advanced LIGO was provided by the Australian Research Council. Virgo is funded, through the European Gravitational Observatory (EGO), by the French Centre National de Recherche Scientifique (CNRS), the Italian Istituto Nazionale di Fisica Nucleare (INFN) and the Dutch Nikhef, with contributions by institutions from Belgium, Germany, Greece, Hungary, Ireland, Japan, Monaco, Poland, Portugal, Spain. KAGRA is supported by Ministry of Education, Culture, Sports, Science and Technology (MEXT), Japan Society for the Promotion of Science (JSPS) in Japan; National Research Foundation (NRF) and Ministry of Science and ICT (MSIT) in Korea; Academia Sinica (AS) and National Science and Technology Council (NSTC) in Taiwan of China.

\emph{\textbf{Statement on the Use of AI}.---}
All scientific ideas, theoretical modeling, calculations, and the original manuscript were developed by the authors. AI-assisted tools such as Codex and BootLoops~\cite{Schwartz2026BootLoops} were used exclusively for language polishing and consistency checks. The authors reviewed and verified all AI-assisted revisions and take full responsibility for the manuscript.

\bibliography{scibib}

@article{Clifton:2011jh,
    author = "Clifton, Timothy and Ferreira, Pedro G. and Padilla, Antonio and Skordis, Constantinos",
    title = "{Modified Gravity and Cosmology}",
    eprint = "1106.2476",
    archivePrefix = "arXiv",
    primaryClass = "astro-ph.CO",
    doi = "10.1016/j.physrep.2012.01.001",
    journal = "Phys. Rept.",
    volume = "513",
    pages = "1--189",
    year = "2012"
}

@article{Maartens:2010ar,
    author = "Maartens, Roy and Koyama, Kazuya",
    title = "{Brane-World Gravity}",
    eprint = "1004.3962",
    archivePrefix = "arXiv",
    primaryClass = "hep-th",
    doi = "10.12942/lrr-2010-5",
    journal = "Living Rev. Rel.",
    volume = "13",
    pages = "5",
    year = "2010"
}

@article{Klein:1926tv,
    author = "Klein, Oskar",
    editor = "Taylor, J. C.",
    title = "{Quantum Theory and Five-Dimensional Theory of Relativity. (In German and English)}",
    doi = "10.1007/BF01397481",
    journal = "Z. Phys.",
    volume = "37",
    pages = "895--906",
    year = "1926"
}

@article{Kaluza:1921tu,
    author = "Kaluza, Th.",
    title = {{Zum Unit{\"a}tsproblem der Physik}},
    eprint = "1803.08616",
    archivePrefix = "arXiv",
    primaryClass = "physics.hist-ph",
    reportNumber = "HUPD-8401",
    doi = "10.1142/S0218271818700017",
    journal = "Sitzungsber. Preuss. Akad. Wiss. Berlin (Math. Phys. )",
    volume = "1921",
    pages = "966--972",
    year = "1921"
}

@article{Finn:1992wt,
    author = "Finn, Lee S.",
    title = "{Detection, measurement and gravitational radiation}",
    eprint = "gr-qc/9209010",
    archivePrefix = "arXiv",
    reportNumber = "PRINT-93-0128 (NORTHWESTERN)",
    doi = "10.1103/PhysRevD.46.5236",
    journal = "Phys. Rev. D",
    volume = "46",
    pages = "5236--5249",
    year = "1992"
}

@article{Biwer:2018osg,
    author = "Biwer, C. M. and Capano, Collin D. and De, Soumi and Cabero, Miriam and Brown, Duncan A. and Nitz, Alexander H. and Raymond, V.",
    title = "{PyCBC Inference: A Python-based parameter estimation toolkit for compact binary coalescence signals}",
    eprint = "1807.10312",
    archivePrefix = "arXiv",
    primaryClass = "astro-ph.IM",
    doi = "10.1088/1538-3873/aaef0b",
    journal = "Publ. Astron. Soc. Pac.",
    volume = "131",
    number = "996",
    pages = "024503",
    year = "2019"
}

@article{Overduin:1997sri,
    author = "Overduin, J. M. and Wesson, P. S.",
    title = "{Kaluza-Klein gravity}",
    eprint = "gr-qc/9805018",
    archivePrefix = "arXiv",
    doi = "10.1016/S0370-1573(96)00046-4",
    journal = "Phys. Rept.",
    volume = "283",
    pages = "303--380",
    year = "1997"
}

@article{Duff:1986hr,
    author = "Duff, M. J. and Nilsson, B. E. W. and Pope, C. N.",
    title = "{Kaluza-Klein Supergravity}",
    doi = "10.1016/0370-1573(86)90163-8",
    journal = "Phys. Rept.",
    volume = "130",
    pages = "1--142",
    year = "1986"
}

@article{Bailin:1987jd,
    author = "Bailin, D. and Love, A.",
    title = "{KALUZA-KLEIN THEORIES}",
    doi = "10.1088/0034-4885/50/9/001",
    journal = "Rept. Prog. Phys.",
    volume = "50",
    pages = "1087--1170",
    year = "1987"
}

@article{ParticleDataGroup:2024cfk,
    author = "Navas, S. and others",
    collaboration = "Particle Data Group",
    title = "{Review of particle physics}",
    doi = "10.1103/PhysRevD.110.030001",
    journal = "Phys. Rev. D",
    volume = "110",
    number = "3",
    pages = "030001",
    year = "2024"
}

@article{Lee:2020zjt,
    author = "Lee, J. G. and Adelberger, E. G. and Cook, T. S. and Fleischer, S. M. and Heckel, B. R.",
    title = "{New Test of the Gravitational $1/r^2$ Law at Separations down to 52 $\mu$m}",
    eprint = "2002.11761",
    archivePrefix = "arXiv",
    primaryClass = "hep-ex",
    doi = "10.1103/PhysRevLett.124.101101",
    journal = "Phys. Rev. Lett.",
    volume = "124",
    number = "10",
    pages = "101101",
    year = "2020"
}

@article{Yang:2012zzb,
    author = "Yang, Shan-Qing and Zhan, Bi-Fu and Wang, Qing-Lan and Shao, Cheng-Gang and Tu, Liang-Cheng and Tan, Wen-Hai and Luo, Jun",
    title = "{Test of the Gravitational Inverse Square Law at Millimeter Ranges}",
    doi = "10.1103/PhysRevLett.108.081101",
    journal = "Phys. Rev. Lett.",
    volume = "108",
    pages = "081101",
    year = "2012"
}

@article{Deffayet:2007kf,
    author = "Deffayet, Cedric and Menou, Kristen",
    title = "{Probing Gravity with Spacetime Sirens}",
    eprint = "0709.0003",
    archivePrefix = "arXiv",
    primaryClass = "astro-ph",
    doi = "10.1086/522931",
    journal = "Astrophys. J. Lett.",
    volume = "668",
    pages = "L143--L146",
    year = "2007"
}

@article{Pardo:2018ipy,
    author = "Pardo, Kris and Fishbach, Maya and Holz, Daniel E. and Spergel, David N.",
    title = "{Limits on the number of spacetime dimensions from GW170817}",
    eprint = "1801.08160",
    archivePrefix = "arXiv",
    primaryClass = "gr-qc",
    doi = "10.1088/1475-7516/2018/07/048",
    journal = "JCAP",
    volume = "07",
    pages = "048",
    year = "2018"
}

@article{MaganaHernandez:2021zyc,
    author = "Magana Hernandez, Ignacio",
    title = "{Constraining the number of spacetime dimensions from GWTC-3 binary black hole mergers}",
    eprint = "2112.07650",
    archivePrefix = "arXiv",
    primaryClass = "astro-ph.HE",
    reportNumber = "LIGO-P2100449",
    doi = "10.1103/PhysRevD.107.084033",
    journal = "Phys. Rev. D",
    volume = "107",
    number = "8",
    pages = "084033",
    year = "2023"
}

@article{Chen:2026owi,
    author = "Chen, Anson and Zhang, Jun",
    title = "{Searching for Extra Dimensions with Gravitational Waves: Dark-Siren Constraints from GWTC-4}",
    eprint = "2606.14549",
    archivePrefix = "arXiv",
    primaryClass = "gr-qc",
    month = "6",
    year = "2026"
}

@article{LIGOScientific:2016aoc,
    author = "Abbott, B. P. and others",
    collaboration = "LIGO Scientific, Virgo",
    title = "{Observation of Gravitational Waves from a Binary Black Hole Merger}",
    eprint = "1602.03837",
    archivePrefix = "arXiv",
    primaryClass = "gr-qc",
    reportNumber = "LIGO-P150914",
    doi = "10.1103/PhysRevLett.116.061102",
    journal = "Phys. Rev. Lett.",
    volume = "116",
    number = "6",
    pages = "061102",
    year = "2016"
}

@article{LIGOScientific:2026sit,
    author = "Abac, None and others",
    collaboration = "LIGO Scientific, VIRGO, KAGRA",
    title = "{GWTC-5.0: An Introduction to Version 5.0 of the Gravitational-Wave Transient Catalog}",
    eprint = "2605.27223",
    archivePrefix = "arXiv",
    primaryClass = "gr-qc",
    reportNumber = "LIGO-P2500701",
    month = "5",
    year = "2026"
}

@article{Qiang:2016chy,
    author = "Qiang, Li-E. and Zhao, Shu Hong and Xu, Peng",
    title = "{Measuring the dimensionality of compact extra dimensions with inspiral gravitational waves from black-hole binaries}",
    doi = "10.1209/0295-5075/116/60001",
    journal = "EPL",
    volume = "116",
    number = "6",
    pages = "60001",
    year = "2016"
}

@article{LIGOScientific:2017vwq,
    author = "Abbott, B. P. and others",
    collaboration = "LIGO Scientific, Virgo",
    title = "{GW170817: Observation of Gravitational Waves from a Binary Neutron Star Inspiral}",
    eprint = "1710.05832",
    archivePrefix = "arXiv",
    primaryClass = "gr-qc",
    reportNumber = "LIGO-P170817",
    doi = "10.1103/PhysRevLett.119.161101",
    journal = "Phys. Rev. Lett.",
    volume = "119",
    number = "16",
    pages = "161101",
    year = "2017"
}

@article{LIGOScientific:2017zic,
    author = "Abbott, B. P. and others",
    collaboration = "LIGO Scientific, Virgo, Fermi-GBM, INTEGRAL",
    title = "{Gravitational Waves and Gamma-rays from a Binary Neutron Star Merger: GW170817 and GRB 170817A}",
    eprint = "1710.05834",
    archivePrefix = "arXiv",
    primaryClass = "astro-ph.HE",
    reportNumber = "LIGO-P1700308",
    doi = "10.3847/2041-8213/aa920c",
    journal = "Astrophys. J. Lett.",
    volume = "848",
    number = "2",
    pages = "L13",
    year = "2017"
}

@article{DES:2017kbs,
    author = "Soares-Santos, M. and others",
    collaboration = "DES, Dark Energy Camera GW-EM",
    title = "{The Electromagnetic Counterpart of the Binary Neutron Star Merger LIGO/Virgo GW170817. I. Discovery of the Optical Counterpart Using the Dark Energy Camera}",
    eprint = "1710.05459",
    archivePrefix = "arXiv",
    primaryClass = "astro-ph.HE",
    reportNumber = "FERMILAB-PUB-17-454-AE-CD-PPD",
    doi = "10.3847/2041-8213/aa9059",
    journal = "Astrophys. J. Lett.",
    volume = "848",
    number = "2",
    pages = "L16",
    year = "2017"
}

@article{Mooley:2022uqa,
    author = "Mooley, Kunal P. and Anderson, Jay and Lu, Wenbin",
    title = "{Optical superluminal motion measurement in the neutron-star merger GW170817}",
    eprint = "2210.06568",
    archivePrefix = "arXiv",
    primaryClass = "astro-ph.HE",
    doi = "10.1038/s41586-022-05145-7",
    journal = "Nature",
    volume = "610",
    number = "7931",
    pages = "273--276",
    year = "2022"
}

@article{Mooley:2018qfh,
    author = "Mooley, K. P. and Deller, A. T. and Gottlieb, O. and Nakar, E. and Hallinan, G. and Bourke, S. and Frail, D. A. and Horesh, A. and Corsi, A. and Hotokezaka, K.",
    title = "{Superluminal motion of a relativistic jet in the neutron-star merger GW170817}",
    eprint = "1806.09693",
    archivePrefix = "arXiv",
    primaryClass = "astro-ph.HE",
    doi = "10.1038/s41586-018-0486-3",
    journal = "Nature",
    volume = "561",
    number = "7723",
    pages = "355--359",
    year = "2018"
}

@article{Mooley:2017enz,
    author = "Mooley, K. P. and others",
    title = "{A mildly relativistic wide-angle outflow in the neutron star merger GW170817}",
    eprint = "1711.11573",
    archivePrefix = "arXiv",
    primaryClass = "astro-ph.HE",
    doi = "10.1038/nature25452",
    journal = "Nature",
    volume = "554",
    pages = "207",
    year = "2018"
}

@article{Cantiello:2018ffy,
    author = "Cantiello, Michele and others",
    title = "{A Precise Distance to the Host Galaxy of the Binary Neutron Star Merger GW170817 Using Surface Brightness Fluctuations}",
    eprint = "1801.06080",
    archivePrefix = "arXiv",
    primaryClass = "astro-ph.GA",
    doi = "10.3847/2041-8213/aaad64",
    journal = "Astrophys. J. Lett.",
    volume = "854",
    number = "2",
    pages = "L31",
    year = "2018"
}

@article{Hu:2017mde,
    author = "Hu, Wen-Rui and Wu, Yue-Liang",
    title = "{The Taiji Program in Space for gravitational wave physics and the nature of gravity}",
    doi = "10.1093/nsr/nwx116",
    journal = "Natl. Sci. Rev.",
    volume = "4",
    number = "5",
    pages = "685--686",
    year = "2017"
}

@article{Du:2025xdq,
    author = "Du, Minghui and others",
    title = "{Towards realistic detection pipelines of Taiji: New challenges in data analysis and high-fidelity simulations of space-based gravitational wave antenna}",
    eprint = "2505.16500",
    archivePrefix = "arXiv",
    primaryClass = "gr-qc",
    doi = "10.1007/s11433-025-2870-8",
    journal = "Sci. China Phys. Mech. Astron.",
    volume = "69",
    number = "4",
    pages = "249501",
    year = "2026"
}

@article{Mangiagli:2022niy,
    author = "Mangiagli, Alberto and Caprini, Chiara and Volonteri, Marta and Marsat, Sylvain and Vergani, Susanna and Tamanini, Nicola and Inchausp{\'e}, Henri",
    title = "{Massive black hole binaries in LISA: Multimessenger prospects and electromagnetic counterparts}",
    eprint = "2207.10678",
    archivePrefix = "arXiv",
    primaryClass = "astro-ph.HE",
    doi = "10.1103/PhysRevD.106.103017",
    journal = "Phys. Rev. D",
    volume = "106",
    number = "10",
    pages = "103017",
    year = "2022"
}

@article{LIGOScientific:2017ync,
    author = "Abbott, B. P. and others",
    collaboration = "LIGO Scientific, Virgo, Fermi GBM, INTEGRAL, IceCube, AstroSat Cadmium Zinc Telluride Imager Team, IPN, Insight-Hxmt, ANTARES, Swift, AGILE Team, 1M2H Team, Dark Energy Camera GW-EM, DES, DLT40, GRAWITA, Fermi-LAT, ATCA, ASKAP, Las Cumbres Observatory Group, OzGrav, DWF (Deeper Wider Faster Program), AST3, CAASTRO, VINROUGE, MASTER, J-GEM, GROWTH, JAGWAR, CaltechNRAO, TTU-NRAO, NuSTAR, Pan-STARRS, MAXI Team, TZAC Consortium, KU, Nordic Optical Telescope, ePESSTO, GROND, Texas Tech University, SALT Group, TOROS, BOOTES, MWA, CALET, IKI-GW Follow-up, H.E.S.S., LOFAR, LWA, HAWC, Pierre Auger, ALMA, Euro VLBI Team, Pi of Sky, Chandra Team at McGill University, DFN, ATLAS Telescopes, High Time Resolution Universe Survey, RIMAS, RATIR, SKA South Africa/MeerKAT",
    title = "{Multi-messenger Observations of a Binary Neutron Star Merger}",
    eprint = "1710.05833",
    archivePrefix = "arXiv",
    primaryClass = "astro-ph.HE",
    reportNumber = "LIGO-P1700294, VIR-0802A-17, FERMILAB-PUB-17-478-A-AE-CD",
    doi = "10.3847/2041-8213/aa91c9",
    journal = "Astrophys. J. Lett.",
    volume = "848",
    number = "2",
    pages = "L12",
    year = "2017"
}

@article{Goldstein:2017mmi,
    author = "Goldstein, A. and others",
    title = "{An Ordinary Short Gamma-Ray Burst with Extraordinary Implications: Fermi-GBM Detection of GRB 170817A}",
    eprint = "1710.05446",
    archivePrefix = "arXiv",
    primaryClass = "astro-ph.HE",
    doi = "10.3847/2041-8213/aa8f41",
    journal = "Astrophys. J. Lett.",
    volume = "848",
    number = "2",
    pages = "L14",
    year = "2017"
}

@article{Pan:2017jem,
    author = "Pan, Y. C. and others",
    title = "{The Old Host-Galaxy Environment of SSS17a, the First Electromagnetic Counterpart to a Gravitational Wave Source}",
    eprint = "1710.05439",
    archivePrefix = "arXiv",
    primaryClass = "astro-ph.GA",
    doi = "10.3847/2041-8213/aa9116",
    journal = "Astrophys. J. Lett.",
    volume = "848",
    number = "2",
    pages = "L30",
    year = "2017"
}

@article{Volonteri:2002vz,
    author = "Volonteri, Marta and Haardt, Francesco and Madau, Piero",
    title = "{The Assembly and merging history of supermassive black holes in hierarchical models of galaxy formation}",
    eprint = "astro-ph/0207276",
    archivePrefix = "arXiv",
    doi = "10.1086/344675",
    journal = "Astrophys. J.",
    volume = "582",
    pages = "559--573",
    year = "2003"
}

@article{Madau:2001sc,
    author = "Madau, Piero and Rees, Martin J.",
    title = "{Massive black holes as Population III remnants}",
    eprint = "astro-ph/0101223",
    archivePrefix = "arXiv",
    doi = "10.1086/319848",
    journal = "Astrophys. J. Lett.",
    volume = "551",
    pages = "L27--L30",
    year = "2001"
}

@article{Volonteri:2007ax,
    author = "Volonteri, M. and Lodato, G. and Natarajan, P.",
    title = "{The evolution of massive black hole seeds}",
    eprint = "0709.0529",
    archivePrefix = "arXiv",
    primaryClass = "astro-ph",
    doi = "10.1111/j.1365-2966.2007.12589.x",
    journal = "Mon. Not. Roy. Astron. Soc.",
    volume = "383",
    pages = "1079",
    year = "2008"
}

@article{Begelman:2006db,
    author = "Begelman, Mitchell C. and Volonteri, Marta and Rees, Martin J.",
    title = "{Formation of supermassive black holes by direct collapse in pregalactic halos}",
    eprint = "astro-ph/0602363",
    archivePrefix = "arXiv",
    doi = "10.1111/j.1365-2966.2006.10467.x",
    journal = "Mon. Not. Roy. Astron. Soc.",
    volume = "370",
    pages = "289--298",
    year = "2006"
}

@article{LSST:2008ijt,
    author = "Ivezi{\'c}, {\v{Z}}eljko and others",
    collaboration = "LSST",
    title = "{LSST: from Science Drivers to Reference Design and Anticipated Data Products}",
    eprint = "0805.2366",
    archivePrefix = "arXiv",
    primaryClass = "astro-ph",
    reportNumber = "SLAC-PUB-16076",
    doi = "10.3847/1538-4357/ab042c",
    journal = "Astrophys. J.",
    volume = "873",
    number = "2",
    pages = "111",
    year = "2019"
}

@inproceedings{Dewdney:2009tmd,
    author = "Dewdney, Peter E. and Hall, Peter J. and Schilizzi, Richard T. and Lazio, T. Joseph L. W.",
    title = "{The Square Kilometre Array}",
    doi = "10.1109/JPROC.2009.2021005",
    month = "8",
    year = "2009"
}

@article{Nandra:2013jka,
    author = "Nandra, Kirpal and others",
    title = "{The Hot and Energetic Universe: A White Paper presenting the science theme motivating the Athena+ mission}",
    eprint = "1306.2307",
    archivePrefix = "arXiv",
    primaryClass = "astro-ph.HE",
    month = "6",
    year = "2013"
}

@article{LISA:2017pwj,
    author = "Amaro-Seoane, Pau and others",
    collaboration = "LISA",
    title = "{Laser Interferometer Space Antenna}",
    journal = {e-prints},
    eprint = "1702.00786",
    archivePrefix = "arXiv",
    primaryClass = "astro-ph.IM",
    month = "2",
    year = "2017"
}

@article{TianQin:2015yph,
    author = "Luo, Jun and others",
    collaboration = "TianQin",
    title = "{TianQin: a space-borne gravitational wave detector}",
    eprint = "1512.02076",
    archivePrefix = "arXiv",
    primaryClass = "astro-ph.IM",
    doi = "10.1088/0264-9381/33/3/035010",
    journal = "Class. Quant. Grav.",
    volume = "33",
    number = "3",
    pages = "035010",
    year = "2016"
}

@article{Padovani:2023dxc,
    author = "Padovani, Paolo and Cirasuolo, Michele",
    title = "{The Extremely Large Telescope}",
    eprint = "2312.04299",
    archivePrefix = "arXiv",
    primaryClass = "astro-ph.IM",
    doi = "10.1080/00107514.2023.2266921",
    journal = "Contemp. Phys.",
    volume = "64",
    number = "1",
    pages = "47--64",
    year = "2023"
}

@misc{GWOSC_GWTC,
  author       = {{LIGO Scientific Collaboration and Virgo Collaboration}},
  title        = {{The Gravitational-wave Transient Catalog}},
  year         = {2025},
  howpublished = {\url{https://gwosc.org/eventapi/html/GWTC/}},
  }

@article{Liu:2023qap,
    author = "Liu, Chang and Ruan, Wen-Hong and Guo, Zong-Kuan",
    title = "{Confusion noise from Galactic binaries for Taiji}",
    eprint = "2301.02821",
    archivePrefix = "arXiv",
    primaryClass = "astro-ph.IM",
    doi = "10.1103/PhysRevD.107.064021",
    journal = "Phys. Rev. D",
    volume = "107",
    number = "6",
    pages = "064021",
    year = "2023"
}

@article{Marsat:2018oam,
    author = "Marsat, Sylvain and Baker, John G.",
    title = "{Fourier-domain modulations and delays of gravitational-wave signals}",
    eprint = "1806.10734",
    archivePrefix = "arXiv",
    primaryClass = "gr-qc",
    month = "6",
    year = "2018"
}

@article{Marsat:2020rtl,
    author = "Marsat, Sylvain and Baker, John G. and Dal Canton, Tito",
    title = "{Exploring the Bayesian parameter estimation of binary black holes with LISA}",
    eprint = "2003.00357",
    archivePrefix = "arXiv",
    primaryClass = "gr-qc",
    doi = "10.1103/PhysRevD.103.083011",
    journal = "Phys. Rev. D",
    volume = "103",
    number = "8",
    pages = "083011",
    year = "2021"
}

@article{Vallisneri:2007ev,
    author = "Vallisneri, Michele",
    title = "{Use and abuse of the Fisher information matrix in the assessment of gravitational-wave parameter-estimation prospects}",
    eprint = "gr-qc/0703086",
    archivePrefix = "arXiv",
    reportNumber = "LIGO-P070009-00-Z",
    doi = "10.1103/PhysRevD.77.042001",
    journal = "Phys. Rev. D",
    volume = "77",
    pages = "042001",
    year = "2008"
}

@article{Cornish:2021lje,
    author = "Cornish, Neil J.",
    title = "{Heterodyned likelihood for rapid gravitational wave parameter inference}",
    eprint = "2109.02728",
    archivePrefix = "arXiv",
    primaryClass = "gr-qc",
    doi = "10.1103/PhysRevD.104.104054",
    journal = "Phys. Rev. D",
    volume = "104",
    number = "10",
    pages = "104054",
    year = "2021"
}

@article{Williams:2021qyt,
    author = "Williams, Michael J. and Veitch, John and Messenger, Chris",
    title = "{Nested sampling with normalizing flows for gravitational-wave inference}",
    eprint = "2102.11056",
    archivePrefix = "arXiv",
    primaryClass = "gr-qc",
    doi = "10.1103/PhysRevD.103.103006",
    journal = "Phys. Rev. D",
    volume = "103",
    number = "10",
    pages = "103006",
    year = "2021"
}

@article{Ashton:2018jfp,
    author = "Ashton, Gregory and others",
    title = "{BILBY: A user-friendly Bayesian inference library for gravitational-wave astronomy}",
    eprint = "1811.02042",
    archivePrefix = "arXiv",
    primaryClass = "astro-ph.IM",
    doi = "10.3847/1538-4365/ab06fc",
    journal = "Astrophys. J. Suppl.",
    volume = "241",
    number = "2",
    pages = "27",
    year = "2019"
}

@article{Nitz:2021zwj,
    author = {Nitz, Alexander H. and Kumar, Sumit and Wang, Yi-Fan and Kastha, Shilpa and Wu, Shichao and Sch{\"a}fer, Marlin and Dhurkunde, Rahul and Capano, Collin D.},
    title = "{4-OGC: Catalog of Gravitational Waves from Compact Binary Mergers}",
    eprint = "2112.06878",
    archivePrefix = "arXiv",
    primaryClass = "astro-ph.HE",
    doi = "10.3847/1538-4357/aca591",
    journal = "Astrophys. J.",
    volume = "946",
    number = "2",
    pages = "59",
    year = "2023"
}

@article{Witten:1981me,
    author = "Witten, Edward",
    title = "{Search for a Realistic Kaluza-Klein Theory*}",
    reportNumber = "PRINT-81-0056 (PRINCETON)",
    doi = "10.1016/0550-3213(81)90021-3",
    journal = "Nucl. Phys. B",
    volume = "186",
    pages = "412",
    year = "1981"
}

@article{Candelas:1985en,
    author = "Candelas, P. and Horowitz, Gary T. and Strominger, Andrew and Witten, Edward",
    title = "{Vacuum configurations for superstrings}",
    reportNumber = "NSF-ITP-84-170",
    doi = "10.1016/0550-3213(85)90602-9",
    journal = "Nucl. Phys. B",
    volume = "258",
    pages = "46--74",
    year = "1985"
}

@article{Eingorn:2010xp,
    author = "Eingorn, Maxim and Zhuk, Alexander",
    title = "{Classical tests of multidimensional gravity: Negative result}",
    eprint = "1003.5690",
    archivePrefix = "arXiv",
    primaryClass = "gr-qc",
    doi = "10.1088/0264-9381/27/20/205014",
    journal = "Class. Quant. Grav.",
    volume = "27",
    pages = "205014",
    year = "2010"
}

@article{Du:2020rlx,
    author = "Du, Yuchen and Tahura, Shammi and Vaman, Diana and Yagi, Kent",
    title = "{Probing Compactified Extra Dimensions with Gravitational Waves}",
    eprint = "2004.03051",
    archivePrefix = "arXiv",
    primaryClass = "gr-qc",
    doi = "10.1103/PhysRevD.103.044031",
    journal = "Phys. Rev. D",
    volume = "103",
    number = "4",
    pages = "044031",
    year = "2021"
}

@article{Chopovsky:2013dg,
    author = "Chopovsky, Alexey and Eingorn, Maxim and Zhuk, Alexander",
    title = "{Many-body problem in Kaluza-Klein models with toroidal compactification}",
    eprint = "1302.0501",
    archivePrefix = "arXiv",
    primaryClass = "gr-qc",
    doi = "10.1140/epjc/s10052-013-2700-7",
    journal = "Eur. Phys. J. C",
    volume = "74",
    pages = "2700",
    year = "2014"
}

@article{Yunes:2009ke,
    author = "Yunes, Nicolas and Pretorius, Frans",
    title = "{Fundamental Theoretical Bias in Gravitational Wave Astrophysics and the Parameterized Post-Einsteinian Framework}",
    eprint = "0909.3328",
    archivePrefix = "arXiv",
    primaryClass = "gr-qc",
    doi = "10.1103/PhysRevD.80.122003",
    journal = "Phys. Rev. D",
    volume = "80",
    pages = "122003",
    year = "2009"
}

@article{Lehners:2022xds,
    author = "Lehners, Jean-Luc",
    title = "{Allowable complex scalars from Kaluza-Klein compactifications and metric rescalings}",
    eprint = "2209.14669",
    archivePrefix = "arXiv",
    primaryClass = "hep-th",
    doi = "10.1103/PhysRevD.107.046004",
    journal = "Phys. Rev. D",
    volume = "107",
    number = "4",
    pages = "046004",
    year = "2023"
}

@article{Cornish:2011ys,
    author = "Cornish, Neil and Sampson, Laura and Yunes, Nicolas and Pretorius, Frans",
    title = "{Gravitational Wave Tests of General Relativity with the Parameterized Post-Einsteinian Framework}",
    eprint = "1105.2088",
    archivePrefix = "arXiv",
    primaryClass = "gr-qc",
    doi = "10.1103/PhysRevD.84.062003",
    journal = "Phys. Rev. D",
    volume = "84",
    pages = "062003",
    year = "2011"
}

@article{Tahura:2018zuq,
    author = "Tahura, Sharaban and Yagi, Kent",
    title = "{Parameterized Post-Einsteinian Gravitational Waveforms in Various Modified Theories of Gravity}",
    eprint = "1809.00259",
    archivePrefix = "arXiv",
    primaryClass = "gr-qc",
    doi = "10.1103/PhysRevD.98.084042",
    journal = "Phys. Rev. D",
    volume = "98",
    number = "8",
    pages = "084042",
    year = "2018",
    note = "[Erratum: Phys.Rev.D 101, 109902 (2020)]"
}

@article{Zhu:2026wtl,
    author = "Zhu, Jie and Song, Hanlin and Lyu, Zhenwei and Li, Hao and Ji, Peixiang and Wang, Jun-Chen and Yan, Haobo and Ma, Bo-Qiang",
    title = "{Testing the strong equivalence principle with multimessenger binary neutron star mergers}",
    eprint = "2603.25413",
    archivePrefix = "arXiv",
    primaryClass = "gr-qc",
    doi = "10.1038/s42005-026-02878-z",
    month = "3",
    year = "2026"
}

@article{Song:2025dbf,
    author = "Song, Hanlin and Li, Hao and Lyu, Zhenwei and Zhu, Jie and Wang, Jun-Chen and Ji, Peixiang",
    title = "{Gravitational-wave constraints on noncommutative spacetime from GW190814}",
    eprint = "2507.23442",
    archivePrefix = "arXiv",
    primaryClass = "gr-qc",
    doi = "10.1103/mcf5-1wvt",
    journal = "Phys. Rev. D",
    volume = "113",
    number = "6",
    pages = "064028",
    year = "2026"
}

@article{Mezzasoma:2022pjb,
    author = "Mezzasoma, Simone and Yunes, Nicol\'as",
    title = "{Theory-agnostic framework for inspiral tests of general relativity with higher-harmonic gravitational waves}",
    eprint = "2203.15934",
    archivePrefix = "arXiv",
    primaryClass = "gr-qc",
    doi = "10.1103/PhysRevD.106.024026",
    journal = "Phys. Rev. D",
    volume = "106",
    number = "2",
    pages = "024026",
    year = "2022"
}

@article{Mehta:2022pcn,
    author = "Mehta, Ajit Kumar and Buonanno, Alessandra and Cotesta, Roberto and Ghosh, Abhirup and Sennett, Noah and Steinhoff, Jan",
    title = "{Tests of general relativity with gravitational-wave observations using a flexible theory-independent method}",
    eprint = "2203.13937",
    archivePrefix = "arXiv",
    primaryClass = "gr-qc",
    reportNumber = "LIGO-P2200083-v2",
    doi = "10.1103/PhysRevD.107.044020",
    journal = "Phys. Rev. D",
    volume = "107",
    number = "4",
    pages = "044020",
    year = "2023"
}

@article{Xie:2024xex,
    author = "Xie, Yiqi and Chung, Adrian Ka-Wai and Sotiriou, Thomas P. and Yunes, Nicol{\'a}s",
    title = "{Bayesian Search of Massive Scalar Fields from LIGO-Virgo-KAGRA Binaries}",
    eprint = "2410.14801",
    archivePrefix = "arXiv",
    primaryClass = "gr-qc",
    doi = "10.1103/PhysRevLett.134.191402",
    journal = "Phys. Rev. Lett.",
    volume = "134",
    number = "19",
    pages = "191402",
    year = "2025"
}

@article{Piarulli:2025rvr,
    author = {Piarulli, Manuel and Marsat, Sylvain and S{\"a}nger, Elise M. and Buonanno, Alessandra and Steinhoff, Jan and Tamanini, Nicola},
    title = "{Parametrized test of general relativity for LISA massive black hole binary inspirals}",
    eprint = "2510.06330",
    archivePrefix = "arXiv",
    primaryClass = "gr-qc",
    doi = "10.1103/59zd-qvbd",
    journal = "Phys. Rev. D",
    volume = "112",
    number = "12",
    pages = "124044",
    year = "2025"
}

@article{McKechan:2010kp,
    author = "McKechan, D. J. A. and Robinson, C. and Sathyaprakash, B. S.",
    editor = "Marka, Zsuzsa and Marka, Szabolcs",
    title = "{A tapering window for time-domain templates and simulated signals in the detection of gravitational waves from coalescing compact binaries}",
    eprint = "1003.2939",
    archivePrefix = "arXiv",
    primaryClass = "gr-qc",
    reportNumber = "LIGO-P0900118-V4",
    doi = "10.1088/0264-9381/27/8/084020",
    journal = "Class. Quant. Grav.",
    volume = "27",
    pages = "084020",
    year = "2010"
}

@article{LISACosmologyWorkingGroup:2026zah,
    author = "Akama, Shingo and others",
    collaboration = "LISA Cosmology Working Group",
    title = "{Testing gravitational wave polarizations with LISA}",
    eprint = "2603.03165",
    archivePrefix = "arXiv",
    primaryClass = "astro-ph.CO",
    doi = "10.1088/1475-7516/2026/09/055",
    journal = "JCAP",
    volume = "09",
    pages = "055",
    year = "2026"
}

@article{Rubbo:2003ap,
    author = "Rubbo, Louis J. and Cornish, Neil J. and Poujade, Olivier",
    title = "{Forward modeling of space borne gravitational wave detectors}",
    eprint = "gr-qc/0311069",
    archivePrefix = "arXiv",
    doi = "10.1103/PhysRevD.69.082003",
    journal = "Phys. Rev. D",
    volume = "69",
    pages = "082003",
    year = "2004"
}

@article{Wu:2019thj,
    author = "Wu, Bofeng and Huang, Chao-Guang and Qiao, Cong-Feng",
    title = "{Analytical analysis on the orbits of Taiji spacecrafts}",
    eprint = "1907.06178",
    archivePrefix = "arXiv",
    primaryClass = "gr-qc",
    doi = "10.1103/PhysRevD.100.122001",
    journal = "Phys. Rev. D",
    volume = "100",
    number = "12",
    pages = "122001",
    year = "2019"
}

@article{Liang:2025izu,
    author = "Liang, Bo and others",
    title = "{Toward Efficient and Accurate EMRI Parameter Estimation: A Machine Learning-Enhanced MCMC Framework}",
    eprint = "2508.00348",
    archivePrefix = "arXiv",
    primaryClass = "gr-qc",
    doi = "10.34133/research.1055",
    journal = "Research",
    volume = "8",
    pages = "1055",
    year = "2026"
}

@article{Liu:2026bbp,
    author = "Liu, Heshan and others",
    title = "{Recent Development of the Laser Interferometer for Taiji Space Gravitational Wave Detection}",
    doi = "10.34133/research.1252",
    journal = "Research",
    volume = "9",
    pages = "1252",
    year = "2026"
}

@article{Tinto:2020fcc,
    author = "Tinto, Massimo and Dhurandhar, Sanjeev V.",
    title = "{Time-delay interferometry}",
    doi = "10.1007/s41114-020-00029-6",
    journal = "Living Rev. Rel.",
    volume = "24",
    number = "1",
    pages = "1",
    year = "2021"
}

@article{Wang:2017aqq,
    author = "Wang, Gang and Ni, Wei-Tou",
    title = "{Numerical simulation of time delay interferometry for TAIJI and new LISA}",
    eprint = "1707.09127",
    archivePrefix = "arXiv",
    primaryClass = "astro-ph.IM",
    doi = "10.1088/1674-4527/19/4/58",
    journal = "Res. Astron. Astrophys.",
    volume = "19",
    number = "4",
    pages = "058",
    year = "2019"
}

@article{Shaddock:2003dj,
    author = "Shaddock, Daniel A. and Tinto, Massimo and Estabrook, Frank B. and Armstrong, J. W.",
    title = "{Data combinations accounting for LISA spacecraft motion}",
    eprint = "gr-qc/0307080",
    archivePrefix = "arXiv",
    doi = "10.1103/PhysRevD.68.061303",
    journal = "Phys. Rev. D",
    volume = "68",
    pages = "061303",
    year = "2003"
}

@article{Tinto:2022zmf,
    author = "Tinto, Massimo and Dhurandhar, Sanjeev and Malakar, Dishari",
    title = "{Second-generation time-delay interferometry}",
    eprint = "2212.05967",
    archivePrefix = "arXiv",
    primaryClass = "gr-qc",
    doi = "10.1103/PhysRevD.107.082001",
    journal = "Phys. Rev. D",
    volume = "107",
    number = "8",
    pages = "082001",
    year = "2023"
}

@article{LIGOScientific:2018hze,
    author = "Abbott, B. P. and others",
    collaboration = "LIGO Scientific, Virgo",
    title = "{Properties of the binary neutron star merger GW170817}",
    eprint = "1805.11579",
    archivePrefix = "arXiv",
    primaryClass = "gr-qc",
    doi = "10.1103/PhysRevX.9.011001",
    journal = "Phys. Rev. X",
    volume = "9",
    number = "1",
    pages = "011001",
    year = "2019"
}

@article{Romero-Shaw:2020owr,
	title        = {\href{https://doi.org/10.1093/mnras/staa2850}{Bayesian inference for compact binary coalescences with bilby: validation and application to the first LIGO\textendash{}Virgo gravitational-wave transient catalogue}},
	author       = {Romero-Shaw, I. M. and others},
	year         = 2020,
	journal      = {Mon. Not. Roy. Astron. Soc.},
	volume       = 499,
	number       = 3,
	pages        = {3295--3319},
	doi          = {10.1093/mnras/staa2850},
	eprint       = {2006.00714},
	archiveprefix = {arXiv},
	primaryclass = {astro-ph.IM}
}

@article{Lin:2023zna,
    author = "Lin, Zi-Chao and Yu, Hao and Gong, Yungui",
    title = "{Limiting the number of extra dimensions with shortcuts}",
    eprint = "2312.17549",
    archivePrefix = "arXiv",
    primaryClass = "gr-qc",
    doi = "10.1103/PhysRevD.109.104015",
    journal = "Phys. Rev. D",
    volume = "109",
    number = "10",
    pages = "104015",
    year = "2024"
}

@article{Wang:2026sir,
    author = "Wang, Hai-Tian and Liang, Dicong and Shao, Lijing and Liu, Yu-Xiao",
    title = "{Testing Extra-Dimensional Gravitational-Wave Polarizations with Compact-Binary Observations}",
    eprint = "2610.11855",
    archivePrefix = "arXiv",
    primaryClass = "gr-qc",
    month = "10",
    year = "2026"
}

@misc{Schwartz2026BootLoops,
  author       = {Schwartz, Matthew D.},
  title        = {{BootLoops 1.0}},
  year         = {2026},
  howpublished = {\url{https://bootloops.ai}},
  doi          = {10.5281/zenodo.23164512}
}

\end{document}